\documentclass[pdflatex,sn-mathphys-num]{sn-jnl}

\usepackage{graphicx}%
\usepackage{multirow}%
\usepackage{amsmath,amssymb,amsfonts}%
\usepackage{amsthm}%
\usepackage{mathrsfs}%
\usepackage[title]{appendix}%
\usepackage{xcolor}%
\usepackage{textcomp}%
\usepackage{manyfoot}%
\usepackage{booktabs}%
\usepackage{algpseudocode}%
\usepackage{listings}%
\usepackage{tikz}
\usetikzlibrary{arrows.meta,calc,decorations.pathreplacing}
\usepackage{algorithm2e}
\usepackage{siunitx}

\theoremstyle{thmstyleone}%
\theoremstyle{thmstyletwo}%

\theoremstyle{thmstylethree}%

\newcommand{\ACBICI}{\texttt{ACBICI}}

\newcommand{\mbs}[1]{{#1}}

\newcommand{\set}[1]{\left\{#1\right\}}

\begin{document}
	\title[Bayesian Calibration]{A Framework for the Bayesian Calibration of Complex and Data-Scarce Models in Applied Sciences}
	
	
	\author*[1]{\fnm{Christina} \sur{Schenk}}\email{christina.schenk@imdea.org, ORCID: 0000-0002-7817-6757}
	
	\author[2,1]{\fnm{Ignacio} \sur{Romero}}\email{ignacio.romero@upm.es, ORCID: 0000-0003-0364-6969}
	
	\affil[1]{ \orgname{IMDEA Materials Institute}, \orgaddress{\street{Calle Eric Kandel 2}, \city{Getafe}, \postcode{28906}, \state{Madrid}, \country{Spain}}}
	
	\affil[2]{ \orgname{Universidad Polit\'ecnica de Madrid}, \orgaddress{\street{Jos\'e Guti\'errez Abascal 2}, \city{Madrid}, \postcode{28006}, \state{Madrid}, \country{Spain}}}

	
	\abstract{In this work, we review the theory involved in the Bayesian calibration of complex computer models, with particular emphasis on their use for applications involving computationally expensive simulations and scarce experimental data. In the article, we present a unified framework that incorporates various Bayesian calibration methods, including well-established approaches. Furthermore, we describe their implementation and use with a new, open-source Python library, \ACBICI (A Configurable BayesIan Calibration and Inference Package). All algorithms are implemented with an object-oriented structure designed to be both easy to use and readily extensible. In particular, single-output and multiple-output calibration are addressed in a consistent manner. The article completes the theory and its implementation with practical recommendations for calibrating the problems of interest. These guidelines -- currently unavailable in a unified form elsewhere -- together with the open-source Python library, are intended to support the reliable calibration of computational codes and models commonly used in engineering and related fields. Overall, this work aims to serve both as a comprehensive review of the statistical foundations and (computational) tools required to perform such calculations, and as a practical guide to Bayesian calibration with modern software tools.}

	
	
	
	\maketitle
	
	\section{Introduction}
	\label{sec-intro}
	Computational models are critical tools in engineering and the applied sciences, providing
	quantitative bases for prediction, understanding, design, and optimization. Their reliability,
	however, depends fundamentally on the proper estimation of uncertain parameters, a process
	traditionally undertaken by minimizing discrepancies between model outputs and experimental
	observations \cite{sung2024review, kennedy2001bayesian,yang_bayesian_2026}. In these classical
	calibration approaches, error minimization by, e.g., least-squares or maximum likelihood, is simple
	to implement, yet leads to well-known challenges including sensitivity to outliers, a reliance on regularization for stability, and limited capacity for uncertainty quantification \cite{schenk2025bayesiancalibrationmodelassessment,DENICOLAO2003669,backman2023}.
	
	The Bayesian calibration framework of Kennedy and O’Hagan (KOH) \cite{kennedy2001bayesian} advanced the field by treating parameter estimation as a statistical inference problem. Within this framework, uncertainties in parameters and model discrepancy are explicitly captured, leading to posterior distributions rather than single point estimates. This framework has been widely adopted across diverse scientific disciplines \cite{higdon2004dl,higdon2008mp,rappel_tutorial_2020,NGUYEN2021,Gattiker2006,OHAGAN2006,bergerson_bayesian_2015,CHONG2018527,Riddle2014,collis_bayesian_2017,de_pablos_experimental_2023,de_pablos_methodology_2020,Beylat2025} and has inspired numerous methodological extensions \cite{spitieris_steinsland_2023,hennicker_etal_2024,sung2024review,EUGENE2023,ling2012calibrationmultiphysicscomputationalmodels,marmin2022,Wilkinson2010}. While most practical implementations to date remain restricted to single-output scenarios, many contemporary engineering systems involve coupled physical phenomena and correlated quantities of interest, making independent single-output calibration statistically inconsistent and potentially misleading \cite{spitieris_steinsland_2023}. Recent research efforts therefore emphasize extensions to multi-output calibration, combining efficient computational strategies and physics-informed priors to extend Bayesian calibration flexibility and scalability in real-world complex systems \cite{spitieris_steinsland_2023}.
	
	Several software packages have been developed to implement the KOH calibration framework. In the R community, CaliCo \cite{carmassi2018calico,calico_docs} is a mature package explicitly developed for Bayesian calibration following the KOH approach. It provides structured, flexible calibration models for both computationally expensive and relatively cheap simulators, with or without model discrepancy. Alongside CaliCo, the Calibrator package \cite{calibrator_2019} offers classical KOH-style Bayesian calibration with strong statistical diagnostics, predominantly applied in environmental and statistical sciences. While these R packages excel in statistical rigor and usability for single-output calibration problems, they tend to be less integrated with machine learning toolkits and high-performance computing resources than newer Python counterparts. Moreover, multi-output calibration remains limited or absent in these tools.
	
	In Python, early work often involved manual implementation of KOH models using probabilistic programming frameworks like PyMC3 \cite{salvatier2016probabilistic}. More recently, dedicated Python packages such as KOH-GPJax \cite{jamesbriant_2023} have emerged, offering scalable Gaussian Process emulators coupled with Hamiltonian Monte Carlo inference tailored for KOH calibration. Additionally, tools like BayCal \cite{idaholab_2022} integrate surrogate modeling with Bayesian calibration workflows designed for computationally demanding simulators, inspired by the KOH methodology but not strictly based on it.
	
	General Bayesian calibration and uncertainty quantification frameworks such as QUEENS \cite{queens_2015, biehler2025queens} offer solver-independent, modular and scalable Python environments supporting advanced sampling and surrogate modeling techniques, while not being strictly based on the KOH approach. Despite these advances, no existing Python library provides a unified, extensible, and user-oriented implementation of both single- and multi-output KOH calibration integrated with diagnostics and practical workflows.
	
	In the current work, we present, in a detailed fashion, the theory behind KOH calibration and we introduce a new
	Python package, \ACBICI~\cite{ACBICIcode,ACBICIdocs}, designed to mirror the key concepts and methods behind this theory
	and extending the formalism to multi-output calibration, enabling simultaneous Bayesian inference
	across multiple outputs. The framework supports modular integration with scientific Python workflows
	and advanced probabilistic programming tools. Key technical features of \ACBICI\ include Gaussian
	process surrogates and efficient algorithms for high-dimensional calibration tasks. Special emphasis
	has been placed on ease-of-use and accessibility for non-Bayesian experts, featuring:
	\begin{itemize}
		\item Intuitive APIs that require minimal prior knowledge in Bayesian statistics,
		\item Comprehensive documentation and tutorials guiding users through calibration workflows,
		\item Built-in default settings that facilitate robust starting points for calibration problems,
		\item Diagnostic plotting and reporting tools designed for clear interpretation,
		\item Support for gradual learning curves, allowing users to progress smoothly from basic calibration to advanced customization.
	\end{itemize}
	Together, these features make the package particularly well-suited for practitioners and domain
	scientists who seek rigorous Bayesian calibration without needing extensive expertise in Bayesian
	statistics or Gaussian process modeling. \ACBICI\ in an object-oriented package whose structure can
	be easily identified with the KOH methods. In this way, researchers that are interested in learning
	and/or extending the KOH methods will find a flexible tool.
	
	Beyond methodological extensions, a main contribution of this work is the inclusion of detailed
	practical recommendations for calibration derived from applied experience and the recent literature. These recommendations include:
	\begin{itemize}
		\item Guidance on robust parameter scaling and normalization for stable inference and convergence.
		\item Strategies for choosing prior distributions.
		\item Use of diagnostic and posterior predictive plots for monitoring calibration quality, analyzing errors, and ensuring meaningful calibration results.
		\item Approaches for assessing posterior sampling algorithms' convergence and mixing, and suggested best practices for efficient sampling.
		\item Global sensitivity analysis for evaluating parameter influence and model robustness.
	\end{itemize}
	These guidelines aim to bridge the gap between theoretical advances and their real-world implementation. By combining methodological innovation with actionable best practices and user-centric design, this package supports rigorous, interpretable, and scalable Bayesian calibration for next-generation computational science.
	
	The main contributions of this article are:
	\begin{enumerate}
		\item A unified Bayesian calibration framework supporting both single-output and multi-output KOH models, summarizing in a unified and self-contained way the KOH theory as well as some related techniques such as sensitivity analysis and sampling.
		\item The description of \ACBICI, a new open-source, object-oriented, extensible Python library implementing these models using
		modern scientific computing practices and following the KOH ideas and expressed in the same form as described in the article.
		\item A consolidated set of practical guidelines for effective Bayesian calibration in data-scarce and computationally demanding settings.
		\item A set of simple examples that hint at the possibilities of \ACBICI, revealing the relationship between the library and the theory, and illustrating the type of outputs that can be obtained from various calibration types.
	\end{enumerate}
	
	The remainder of the article is organized as follows. Section~\ref{sec-background} provides a self-contained account of the KOH approach to model calibration. It describes the calibration strategies implemented in \ACBICI\, including those involving surrogate models, discrepancy functions, experimental error, and multi-output formulations. Section~\ref{sec:sensana} describes the global sensitivity analysis procedures included in \ACBICI. These are, strictly speaking, unrelated to Bayesian calibration, yet they are often used prior to the latter, because they can identify which are the model variables that need more effort in the calibration. Section~\ref{sec-posterior} reviews the two methods available in \ACBICI\ for the computation of the \emph{a posteriori} joint probability distribution of the model parameters. This is a key step in any Bayesian inference task and indisputably the most time-consuming. Our library does not implement any new method for this task, but we include for completeness the essential information about the theory of the libraries that we use, together with links and references to further information. The section concludes (in Section \ref{sec:pred}) by demonstrating how the calibrated parameters can be employed for making predictions using the Gaussian process. Section~\ref{sec-practical} gathers a short list of key recommendations that we found useful when performing model calibration and, furthermore, are often overlooked and not documented. Finally, Section~\ref{sec-examples} illustrates the use of \ACBICI\ for a series of selected examples that illustrate the usage of the library and its provided output. The article is closed by Section~\ref{sec-conclusions}, which summarizes some of the main ideas of the article and collects some final remarks.
	
	\section{Bayesian calibration and prediction}
	\label{sec-background}
	
	\begin{figure}[htb]
		\centering
		\includegraphics[width=0.8\linewidth]{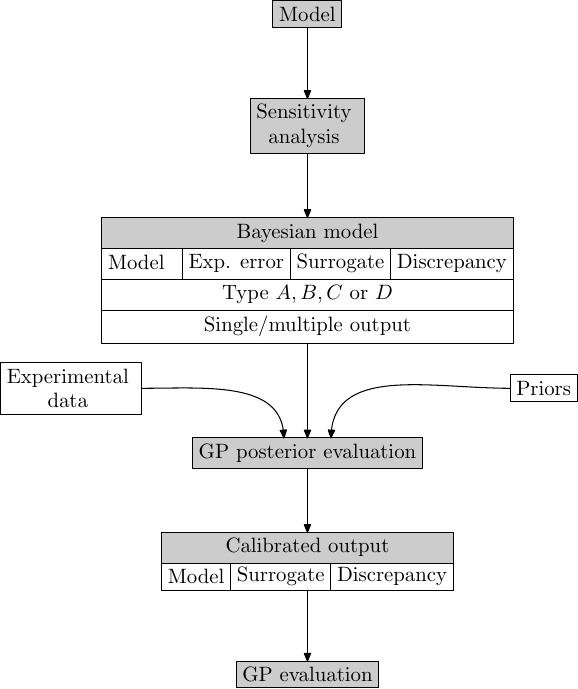}
		\caption{Structure of a Bayesian calibration procedure.}
		\label{fig-structure}
	\end{figure}
	
	The calibration methods discussed in the current article are all based on the seminal work by Kennedy and O'Hagan \cite{kennedy2001bayesian}, later elaborated by several authors (e.~g.,
	\cite{higdon2004dl,williams2006qa,van2005kg,higdon2008mp,mcclarren2018yq,damblin2018iu,schoot2021vr}).
	In this section, we summarize this framework noting, from the outset, that it is fairly general and
	it can encompass several types of calibration problems depending on the features that are included,
	as illustrated in Figure~\ref{fig-structure}. 
	
	Three aspects, which we now identify, may or may not
	be incorporated into the calibration, leading to $2^3\equiv8$ analysis types. First, the model to be
	calibrated might be expensive or inexpensive to run. If of the first type, it will be necessary to
	replace it with a (fast) \emph{surrogate} model that can be evaluated thousands of times to obtain
	statistical significance. Second, if the model is too simple or unsuitable to represent the problem it is supposed to, it might prove necessary to calibrate a certain \emph{discrepancy} function. Third, the experimental data collected to calibrate the model always has some measurement error which might, or might not, need to be calibrated as well. Similar to seen in previous works \cite{carmassi2018calico}, we define eight different calibration methods, summarized in Table~\ref{tab-types}.
	\begin{table}[h]
		\centering
		\begin{tabular}{c c c c}
			Type & Discrepancy & Surrogate & Experimental error \\
			\hline
			$A$ & No & No & Yes/No \\
			$B$ & No & Yes & Yes/No  \\
			$C$ & Yes & No & Yes/No  \\
			$D$ & Yes & Yes & Yes/No  \\
			\hline
		\end{tabular}
		\caption{Calibration types. Each of the four may include the calibration of the variance in the experimental
			error.}
		\label{tab-types}
	\end{table}
	
	Type $A$ calibration is the simplest one, a prototypical example of Bayesian inference where, given
	some prior probability distributions for the parameters, data is employed to learn and improve this
	distribution. Type $B$ is tailored to models that are expensive to run and thus, few model
	evaluations can be made during the calibration. Type $C$ calibration is designed to provide
	information about the adequacy of the model to predict the experimental data, or the contrary. When
	this information is sought and also the model is expensive to run, type $D$ should be used. In all
	cases, the experimental measurements are assumed to be afflicted by some Gaussian error with zero
	mean and whose variance can be estimated, if desired.
	
	\subsection{Sensitivity analysis}
	\label{sec:sensana}
	Often, models depend on a large number of parameters and it proves impractical to calibrate all of
	them, especially if the output is not too sensible to the value of some of these parameters.
	Sensitivity analysis enables the quantification of the influence of uncertain model parameters on
	the variability of model outputs. Performing this analysis prior to calibration is recommended,
	particularly when dealing with models that depend on a large number of parameters or when challenges
	in estimating parameters are encountered. The sensitivity analysis is crucial for identifying the
	most influential parameters, guiding model refinement, and reducing the dimensionality of
	calibration tasks.
	
	To address these aspects, \ACBICI\ provides tools for global sensitivity analysis. Among the
	different approaches available for these analyses, we have implemented variance-based global
	sensitivity analysis using Sobol indices \cite{sobol1993}. Such a method decomposes the variance of the model output into contributions from each parameter and their interactions, thus offering a complete picture of parameter importance.
	
	To explain the basic ideas of global sensitivity analysis, consider a model $Y = f(Q)$ that depends
	on $p$ random variables $Q=(Q_1, Q_2, \ldots, Q_p)\in\mathbb{R}^{p}$ and $Y\in \mathbb{R}$. The total variance of the random output can be expressed as
	\begin{equation}
		\mathrm{Var}[Y] = \sum_{i=1}^{p} V_i + \sum_{1\le i<j\le p} V_{ij}\ ,
	\end{equation}
	where $V_i = \mathrm{Var}[\mathbb{E}[Y|Q_i]]$ denotes the contribution of the individual parameter
	$Q_i$ with $\mathbb{E}$ being the expectation, and higher-order terms represent contributions from parameter interactions.
	Two sets of Sobol indices are typically computed. First, the \emph{first-order Sobol index} of parameter $Q_i$ is defined as:
	\begin{equation}
		S_i = \frac{V_i}{\mathrm{Var}[Y]},
		\quad
		i=1,\ldots,p,
	\end{equation}
	apportions the variance attributable to $Q_i$ alone. Second, the \emph{total-order Sobol index}
	corresponding to parameter $Q_i, 1\le i\le p$ is defined as
	\begin{equation}
		S_{T_i} = \frac{V_i}{\mathrm{Var}[Y]} + \sum_{i=1}^{p} \frac{V_{ij}}{\mathrm{Var}[Y]}.
	\end{equation}
	One can show that the total sensitivity index can be interpreted as (e.g.~\cite{smith2024})
	\begin{equation}
		S_{T_i} = 1 - \frac{\mathrm{Var}[\mathbb{E}[Y|Q_{\sim i}]]}{\mathrm{Var}[Y]}= \frac{\mathbb{E}[\mathrm{Var}[Y|Q_{\sim i}]]}{\mathrm{Var}[Y]},
	\end{equation}
	where $Q_{\sim i}$ indicates the set of all parameters except $Q_i$. This index quantifies the overall contribution of $Q_i$, including both its main and interaction effects.
	
	Random samples of the parameter space are generated using quasi-Monte Carlo techniques, and variance-based estimators are used to compute both first-order and total-order indices. The results allow practitioners to rank parameters, identify negligible ones, and focus the calibration process on the most influential quantities. This functionality is particularly valuable when high-dimensional problems are considered, as it can reduce computational burden by guiding model simplification. 
	
	\ACBICI\ provides Sobol analysis leveraging the \texttt{SALib} Python package. 
	
	Future work could include an estimability analysis, for example based on orthogonalization \cite{yao2003modeling,Shaohua2011,KIPET2020}, to complement the global sensitivity analysis and provide further insight into parameter identifiability and inference robustness.
	
	\subsection{Case A: simple calibration}
	\label{subs-caseA}
	This analysis type is employed to calibrate \emph{inexpensive} models that are assumed to be \emph{adequate}
	enough to reproduce the available experimental data. The starting \emph{ansatz} of type $A$
	calibration is that the measurement $y\in \mathbb{R}$ obtained from running an experiment with controlled variable
	$x\in \mathbb{R}^{d}$ is related to the model $m:\mathbb{R}^d\times \mathbb{R}^p\to \mathbb{R}$
	through the identity
	\begin{equation}
		\label{eq-typeA-ansatz}
		y = m(x;\theta) + \epsilon(x)\ .
	\end{equation}
	Here $\theta\in \mathbb{R}^p$ is a vector of model parameters and $\epsilon$ is a random variable that
	represents the noise in the measurements, distributed according to a Gaussian with zero mean and
	variance~$\sigma^2$.
	
	In every instance, a Bayesian calibration requires three ingredients: a \emph{prior} probability
	distribution for the unknown parameters, data, and a likelihood function. Assuming first that the
	variance of the experimental data is known, for type $A$ calibration, only the prior for the
	parameters $\pi(\theta)$ is required. To proceed, the experimental data are collected in a set
	\begin{equation}
		\label{eq-data}
		\mathcal{D} = \set{ (x_i,y_i)\in \mathbb{R}^d\times \mathbb{R}^p, i=1,2,\ldots N}\ ,
	\end{equation}
	and the likelihood is obtained from the distribution of the experimental error to be
	\begin{equation}
		\label{eq-typea-like}
		L(\mathcal{D} \vert \theta) =
		\frac{1}{(2\pi)^{N/2} \sigma^{N}}
		\exp\left[-\frac{1}{2\,\sigma^2} \sum_{i=1}^N |y_{i}-m(x_i;\theta)|^2\right]\ .
	\end{equation}
	Then, by Bayes' theorem, using the previous information, the posterior probability is simply
	\begin{equation}
		\label{eq-typea-posterior}
		p(\theta \vert \mathcal{D}) \propto
		L(\mathcal{D} \vert \theta)\;\pi(\theta)\ .
	\end{equation}
	For arbitrary priors, the evaluation of this posterior distribution cannot be performed analytically
	and requires using a
	numerical method such as the Markov Chain Monte Carlo (MCMC) method (e.g.,
	\cite{madras2002uh,robert2004gp,barbu2020is,romero2025qk}). See Section~\ref{sec-posterior} for
	details and alternative approaches.
	
	If the variance of the experimental error is unknown, the starting prior must be the joint
	probability distribution $\pi(\theta,\sigma)$. In this case, the functional form of the likelihood~\eqref{eq-typea-like}
	remains unchanged, but it should now be considered a function of $\sigma$ as well.
	Also, the posterior distribution~\eqref{eq-typea-posterior} must also be corrected to
	\begin{equation}
		\label{eq-posteriora-exp}
		p(\theta,\sigma  \vert \mathcal{D}) \propto
		L(\mathcal{D} \vert \theta, \sigma)\; \pi(\theta, \sigma)\ .
	\end{equation}

	\subsection{Case B: calibration of expensive models}
	\label{subs-caseb}
	
	The second relevant class of calibration problems deals with situations where the model to be
	calibrated is too expensive to be evaluated many times --- as required by the MCMC algorithm --- or
	the extreme case where the experimental data consists of some legacy model evaluations but it cannot
	be evaluated any more. This happens often when the model is a complex computer simulation such as a
	finite element discretization.
	
	In cases such as these, the calibration depends on two types of data:
	\begin{itemize}
		\item A collection of \emph{experimental} datapoints $\{\bar{x}_i,\bar{y}_i \}_{i=1}^N$.
		\item A collection of \emph{synthetic} datapoints $\{\tilde{x}_i, \tilde{t}_{i}, \tilde{y}_i \}_{i=1}^M$ obtained by evaluating the (expensive) model with input variables and parameters $(\tilde{x}_i,\tilde{t}_i)$, and whose result is
		$\tilde{y}_i= m(\tilde{x}_i,\tilde{t}_{i})$. The assumption of type $B$ calibration is that this set
		of synthetic points can not be enlarged with additional data, and alternatives avenues for
		evaluating --- at least \emph{approximately} --- the model should be found.
	\end{itemize}
	
	For convenience, we gather all the input variables into a single array
	${x}=\{\bar{x}_i,\tilde{x}_j\}$ of dimension $N+M$, the outputs also into an array
	$y=\{ \bar{y}_i,\tilde{y}_j\}$ of dimension, $N+M$, and define the array ${t} = \{
	\tilde{t}_i \}_{i=1}^M$. In this way, all the available data, be it experimental or synthetic, can
	be collected into
	\begin{equation}
		\label{eq-data-b}
		\mathcal{D} =  \left\{ x,t,y \right\}.
	\end{equation}

	When calibrating expensive models it is necessary to build an \emph{emulator} or \emph{surrogate}
	model that will replace the original one in the evaluations required to calculate the posterior
	probability; see Section \ref{sec-posterior}. The goal is to construct a \emph{fast}
	method of evaluating \emph{approximately} the original model. Naturally, some information of the
	complex model will be lost in this approximation process or, in other words, more uncertainty will
	enter the calibration. This is clearly not a desirable feature, but in situations like the ones
	under consideration it might be unavoidable. Accepting this limitation, the relation~\eqref{eq-typeA-ansatz} must be replaced with
	\begin{equation}
		\label{eq-ansatz-B}
		y = \tilde{m}(x;\theta) + \epsilon(x)\ ,
	\end{equation}
	where $\tilde{m}$ is now the emulator for model~$m$.
	
	Surrogate models ---also referred to as \emph{meta-models}--- can be, in principle, any type of fast
	regression scheme (neural networks \cite{hastie2009vr}, random forest models \cite{breiman2001kr}, even simple
	analytical models if they can be found). However, when using a surrogate, it is not
	easy to account for the uncertainty in its predictions or, more precisely, to integrate this
	uncertainty into the general calibration framework. Gaussian processes (GPs) \cite{rasmussen2006vz}, by contrast, provide accurate emulators even for relatively scarce data and integrate naturally into the Bayesian structure advocated here. Moreover, these surrogates
	have an intrinsic measure of uncertainty that can be combined with the other sources in the
	calibration.
	
	A review of GPs is beyond the goals of this article, and we refer to standard
	references for their detailed description \cite{rasmussen2006vz} and to Section~\ref{sec:pred} for
	their use as predictors. For the moment, we will only recall
	that they are uniquely defined in terms of a mean and a covariance function denoted as $\mu(x,t)$
	and $c_S(x,t; x',t')$ (the symbol $S$ refers to the \emph{surrogate}) evaluated at pairs of input variables and parameters, and that depends on
	hyperparameters collectively referred to as $\chi$. For example, the mean function is usually taken
	to be identically zero and the covariance is often built using a kernel function of the distance between two pairs
	$(x,t)$ and $(x',t')$. The parameters of this kernel are precisely the hyperparameters $\chi$.
	
	The Bayesian calibration of expensive models requires, as always, data, priors for the
	unknown probability distributions, and a likelihood function. The data in this case is defined in Eq.~\eqref{eq-data-b}, and includes experimental and synthetic data. The prior
	distribution must refer to the unknown parameters $\theta$, the variance $\sigma^2$ of the
	experimental error (if unknown), and the hyperparameters $\chi$ of the surrogate; we indicate this
	prior as $\pi(\theta,\sigma,\chi)$, and we note that it is usually defined as the product of three independent
	density probabilities. 
	
	Finally, assuming that the surrogate is a GP, the likelihood of the data~$\mathcal{D}$, given
	some values for $\theta,\chi,\sigma$, is
	\begin{equation}
		\label{eq-likelihood-B}
		L(\mathcal{D} \vert \theta,\chi,\nu)
		\propto
		\det(\Sigma)^{-1/2}
		\exp
		\left[
		\frac{-1}{2} {y}\cdot \Sigma^{-1} {y}
		\right]
	\end{equation}
	with $\Sigma$ being the covariance matrix
	\begin{equation}
		\label{eq-covariance-B}
		\Sigma
		=
		\begin{bmatrix}
			\mbs{C}_{11}  & \mbs{C}_{12} \\
			\mbs{C}_{21} & \mbs{C}_{22} \\
		\end{bmatrix},
	\end{equation}
	where $\mbs{C}_{11},\mbs{C}_{12}=\mbs{C}_{21}^T$ and $\mbs{C}_{22}$ are blocks with components
	\begin{equation}
		\label{eq-blocks-A}
		\begin{aligned}
			C_{11}^{ij} &= c_S( \bar{x}_i,\theta ;  \bar{x}_j,\theta)+ \sigma^2\delta_{ij}\ , \\
			C_{12}^{ik} &= c_S( \bar{x}_i,\theta ;          \tilde{x}_k,\tilde{t}_{k}) \ , \\
			C_{22}^{kl} &= c_S( \tilde{x}_k,\tilde{t}_{k} ; \tilde{x}_l,\tilde{t}_{l})\ ,
		\end{aligned}
	\end{equation}
	$\delta_{ij}$ being the Kronecker delta, and $1 \le i,j \le N$, $1 \le k,l \le M$. Combining the prior distributions with the
	likelihood~\eqref{eq-likelihood-B}, Bayes' theorem provides the posterior probability distribution
	\begin{equation}
		\label{eq-posteriorB}
		p(\theta, \sigma, \chi \vert \mathcal{D})
		\propto
		L(\mathcal{D} \vert \theta,\chi,\nu)\;
		\pi (\theta,\chi,\nu)\ .
	\end{equation}
	Note that the calibration of expensive models will resort to the meta-model to evaluate $M$ points
	$\tilde{y}_i = \tilde{m}(\tilde{x}_i,\tilde{t}_i)$ for each point in the $MCMC$ solution. Given that
	$M\gg N$, and that the number of MCMC evaluations is typically in the order of thousands, it pays
	off to work with a fast accurate meta-model.
	
	\subsection{\ACBICI\ surrogate models} 
	The standard surrogate models implemented in \ACBICI\ are based on GPs. These models rely on kernel functions to encode similarity between input points, making the notion of distance fundamental to their construction. The covariance function is defined through a kernel $k:\mathbb{R}^+\cup\{0\} \to \mathbb{R}\cup \{0\}$ such that
	\begin{equation}
		\label{eq-covariance-kernel}
		c(\mbs{x},\mbs{y}) \;=\; k\!\left(\|\mbs{x}-\mbs{y}\|\right)\;,
	\end{equation}
	where $\|\cdot\|$ denotes the Euclidean norm. 
	In \ACBICI, all variables are assumed to be dimensionless, drastically simplifying the kernel
	construction, yet introducing limitations. First, physical interpretability may be reduced, since dimension information is lost and relationships that depend on physical scales become obscured. Second, forcing heterogeneous variables into a common nondimensional representation may restrict applicability in multi-physics or multi-scale settings. Third, without an explicit metric, kernel-based methods become sensitive to arbitrary rescaling of the inputs, potentially biasing the GP's similarity structure.
	
	Under the dimensionless-variable assumption, \ACBICI\ employs standard isotropic kernels that depend only on the scalar distance $r=\|\mbs{x}-\mbs{x}'\|$. Common choices include the squared exponential (RBF - Radial Basis Function) kernel:
	\begin{equation}
		\label{eq-rbf}
		k_{\mathrm{sqexp}}(r;\lambda,\beta)
		= \lambda \exp\!\left[-\frac{r^2}{2\beta^2}\right],
	\end{equation}
	along other options such as the Matérn $3/2$ kernel,
	the Matérn $5/2$ kernel,
	the exponential kernel,
	and the rational quadratic kernel with $\alpha=1$.
	For multi-output GPs, \ACBICI\ adopts a similarity kernel that uses the Matérn $3/2$ kernel for
	comparisons within the same task and assigns zero covariance across different tasks; see Section~\ref{sec:multicalib}.
	
	All kernels contain one or more hyperparameters that determine their behavior, most importantly the
	signal variance $\lambda$ and the characteristic lengthscale $\beta$. In \ACBICI, separate
	lengthscale parameters are employed for model parameters (indicated later as $\beta_t$) and for
	input variables (denoted as $\beta_x$), allowing independent control over smoothness in each space.
	\subsection{Case C: calibration of models with discrepancy}
	\label{subs-casec}
	The calibration approaches described in Sections~\ref{subs-caseA} and \ref{subs-caseb} assume, as
	starting hypothesis, that the model to be calibrated is able to
	accurately approximate the experimental results, when corrected by the experimental error. This is
	a strong assumption that is not always satisfied. It is extremely common, in fact, that
	simple models are favored over complex ones, at the expense of having poor accuracy or, in the worst
	case, unforeseen bias. Case $C$ calibration starts from a different \emph{ansatz}: the selected model
	might be insufficiently accurate to represent the phenomenon it is supposed to, and, hence, there is
	some \emph{discrepancy} between the measurements and the model predictions that can not be fully
	removed even by selecting optimal parameter distributions.
	
	Assuming that there exists some discrepancy between the model and the underlying phenomenon, we can
	express the relationship between the measurements $y\in \mathbb{R}$, the input $x\in \mathbb{R}^d$,
	and the parameters $\theta\in \mathbb{R}^p$ as
	\begin{equation}
		\label{eq-ansatz-C}
		y = m(x;\theta) + \delta(x) + \epsilon(x)\ .
	\end{equation}
	In this expression, $m:\mathbb{R}^d\times \mathbb{R}^p\to \mathbb{R}$ is the model to be calibrated,
	$\delta$ is a discrepancy function, and $\epsilon$ is, as before, the measurement error. In type $C$,
	we strive to calibrate the model $m$ by finding the optimal probability distributions for the
	parameters $\theta$ and $\sigma^{2}$, while simultaneously searching for the discrepancy $\delta$
	that helps to explain best the data. In contrast with case $B$, in type $C$ calibration we do not take
	the model $m$ to be expensive to evaluate.
	
	The discrepancy function might be selected among a class of trial functions. As we mentioned with
	the choice of surrogate of type $B$ calibration, the best discrepancy function would be \emph{form
		free}, meaning that its shape should not be constrained prior to the calibration. Additionally, it is desirable
	that modeling the discrepancy does not interfere with the other calibrations of the process. Again,
	GPs possess all the required features and thus are favored in the calibration framework described
	herein.

	Replicating the same notation as in previous calibration types, let $\mathcal{D}=\{(x_i,y_i)\}_{i=1}^N \equiv (\mbs{x},\mbs{y})\}$
	be the set of experimental input/output pairs. If the discrepancy $\delta$ is modeled with a
	GP of zero mean, covariance function $c_\delta(x,x')$ depending on hyperparameters $\psi$,
	then the likelihood of $\mathcal{D}$, given $\theta$ and $\psi$ is
	\begin{equation}
		\label{eq-likelihoodD}
		L( \mathcal{D} \vert \theta,\psi,\sigma)
		\propto
		\det(\Sigma)^{-1/2}
		\exp
		\left[
		-\frac{1}{2}
		(\mbs{y}-m(\mbs{x},\theta))\cdot
		\Sigma^{-1}
		(\mbs{y}-m(\mbs{x},\theta))
		\right],
	\end{equation}
	where the covariance matrix $\Sigma$ has components
	\begin{equation*}
		\Sigma_{ij} = c_\delta(x_i,x_j) + \sigma^2\,\delta_{ij}\ ,
	\end{equation*}
	and $1\le i,j \le N$. As in all previous
	calibrations, the posterior probability for the unknown parameters, hyperparameters, and variance of
	the experimental error can be obtained from Bayes' rule which, in case $C$, takes the form
	\begin{equation}
		\label{eq-typeC-bayes}
		p(\theta,\psi,\sigma \vert \mathcal{D}) \propto
		L( \mathcal{D} \vert \theta,\psi,\sigma) \; \pi(\theta,\psi,\sigma)\ .
	\end{equation}
	Here, $\pi(\theta,\psi,\sigma)$ is the joint prior probability distribution of the unknown random
	variables, namely, the model parameters, the hyperparameters of the discrepancy covariance function,
	and the standard deviation of the experimental error, if unknown.

	\subsection{Case D: calibration of expensive models with discrepancy}
	\label{subs-cased}
	Finally, we might want to combine the features of cases $B$ and $C$ into one calibration scheme that
	makes use of a surrogate for the model and incorporates a discrepancy function too.
	
	The steps to
	perform this calibration are a combination of the ones employed in previous scenarios. As in case $B$,
	let $\mathcal{D}=\{(x,t,y)\}$ collect the $N$ experimental input/output pairs and the $M$ synthetic data generated with the original (expensive) model, just as in Eq.~\eqref{eq-data-b}. We would like to model the experimental results $y$ starting from the hypothesis
	\begin{equation}
		\label{eq-ansatzD}
		y = \tilde{m}(x;\theta) + \delta(x) + \epsilon(x)\ ,
	\end{equation}
	where $\tilde{m}$ is the model surrogate, $\delta$ is the discrepancy function, and $\epsilon$ is
	the experimental error. Next, we assume that the meta-model is a GP with zero mean,
	covariance $c_S(x,t;x',t')$, and hyperparameters $\chi$. Similarly, the discrepancy function is taken
	to be another GP with zero mean function, covariance function $c_{\delta}(x,x')$, and
	hyperparameters~$\psi$. Finally, the experimental error is assumed to be normally distributed with
	zero mean and variance~$\sigma^2$. To set up the likelihood of the data, let $y$ be the $N+M$ dimensional vector collecting the experimental measurements and synthetic outputs and define
	\begin{equation}
		\label{eq-caseD-likelihood}
		L(\mathcal{D} \vert \theta,\chi,\psi,\sigma)
		\propto
		\det(\Sigma)^{-1/2}
		\exp
		\left[
		\frac{-1}{2} \mbs{y}\cdot \Sigma^{-1} \mbs{y}
		\right]\ ,
	\end{equation}
	with a covariance matrix $\Sigma$ that has the following block structure:
	\begin{equation}
		\label{eq-caseD-block-covariance}
		\Sigma
		=
		\begin{bmatrix}
			\mbs{C}_{11} & \mbs{C}_{12} \\ \mbs{C}_{21} & \mbs{C}_{22}
		\end{bmatrix}\ ,
	\end{equation}
	with $\mbs{C}_{12}=\mbs{C}_{21}^T$. The entries of these blocks are now
	\begin{equation}
		\label{eq-caseD-entries-sigma}
		\begin{aligned}
			C_{11}^{ij} &= c_S(x_i,\theta;x_j,\theta) + c_{\delta}(x_i,x_j) + \sigma^2 \delta_{ij}\ , \\
			C_{12}^{ik} &= c_S(x_i,\theta; x_k, t_k)\ , \\
			C_{22}^{kl} &= c_S(x_k,t_k; x_l, t_l)\ ,
		\end{aligned}
	\end{equation}
	with $1\le i,j \le N$ and $1\le k,l \le M$. If the prior $\pi(\theta,\chi,\psi,\sigma)$ of the
	unknown random variables is given, their joint posterior distribution is given again by Bayes' theorem through the expression:
	\begin{equation}\label{eq-typeD-bayes}
		p(\theta,\chi,\psi,\sigma \vert \mathcal{D}) \propto
		L(\mathcal{D} \vert \theta,\chi,\psi,\sigma) \; \pi(\theta,\chi,\psi,\sigma)\ ,
	\end{equation}
	whose evaluation must rely on an algorithm such as MCMC.

	\subsection{Synthetic data generation in practice}
	The use of surrogate models in cases~$B$ and~$D$ is motivated by the fact that the original forward
	model is prohibitively expensive to evaluate the hundreds to tens of thousands of times required by
	standard MCMC sampling procedures. Instead, a set of carefully selected points is generated to cover
	the input and parameter spaces using, for example, a Latin Hypercube sampler. To construct the
	surrogate, the original model is then evaluated only at these selected points.
	
	However, when the dimension of the input space is high, the synthetic points may lie far from the values observed in the experiments. Moreover, and especially when the parameter space is also high-dimensional, many of the sampled points may correspond to model outputs that deviate substantially from the experimental measurements. These observations imply that a large number of synthetic points would be needed for the surrogate to faithfully approximate the original model, particularly in regions near the experimental data. Since each synthetic evaluation introduces computational overhead, it is crucial to train the surrogate using points that are as informative as possible while still ensuring adequate coverage of the full input space.
	
	As in many areas of data science, an effective strategy must balance \emph{exploitation}, sampling synthetic points near the experimental data to improve local surrogate accuracy, with \emph{exploration}, sampling throughout the input space to avoid bias and ensure global representativeness. 
	
	In the examples included in this version of \ACBICI, we employ Latin Hypercube
	Sampling (LHS) to generate synthetic data points.
	To ensure that this dataset is representative of the underlying model, the input
	and parameter spaces must be sufficiently covered. While LHS provides an efficient
	space-filling design and is well suited for moderate-dimensional problems, it may
	become increasingly costly and less effective as the dimensionality grows \cite{mckay1979xs,iman1981nxy}. In
	particular, although LHS is designed to avoid clustering and promote uniform
	coverage along each dimension, clustering effects can still arise in
	high-dimensional settings and are further exacerbated in the presence of
	constraints~\cite{Esposito2023,schenk2024castroefficientconstrained}. For more complex or high-dimensional applications, alternative design-of-experiment or adaptive sampling strategies, such as active learning approaches, should be employed \cite{schenk2024castroefficientconstrained, bashiri_comparison_2022,kamath_intelligent_2022}.
	
	\subsection{Multi-output calibration}
	\label{sec:multicalib}
	It is not uncommon that a single system --- be it chemical, physical, social, or otherwise ---
	provides interrelated outputs that are modeled with codes or equations which share parameters. When
	experimental data is available for these distinct outputs, and especially if it is scarce, it makes
	sense to combine the information coming from different --- yet connected --- experimental data and
	perform a joint, \emph{multi-output} calibration. In the context of Bayesian methods, multi-output
	models are much more difficult to calibrate than single-output ones and therefore are often not
	addressed. Here, we describe a way to extend the KOH framework to multi-output problems. This is an
	advanced feature of \ACBICI\ that is not available in existing Bayesian calibration tools.
	
	To extend the Bayesian calibration framework described heretofore to
	multi-output scenarios, we augment the input space with a task index,
	thereby reducing the multi-output case to an equivalent single-output
	formulation in an enlarged input space. Specifically, the augmented input space is defined as
	\begin{equation}
		(x,i) = r \in \mathbb{R}^d \times \{0,\ldots,n_{\text{tasks}}-1\},
		\label{eq:augmented-input}
	\end{equation}
	where $n_{\text{tasks}}$ denotes the number of calibration tasks or outputs.
	
	We consider task-specific models
	$m_i : \mathbb{R}^d \times \mathbb{R}^p \to \mathbb{R}$,
	for $i = 0,\ldots,n_{\text{tasks}}-1$ that are functions of the same input
	and parameter spaces. Then, we define an effective single-output model
	\[
	m : \mathbb{R}^d \times \{0,\ldots,n_{\text{tasks}}-1\} \times \mathbb{R}^p
	\to \mathbb{R}
	\]
	as
	\begin{equation}
		\label{eq-combined-model}
		m(x,i,\theta)
		=
		\sum_{j=0}^{n_{\text{tasks}}-1} m_j(x,\theta)\,\delta_{ij},
	\end{equation}
	where $\delta_{ij}$ denotes the Kronecker delta. This construction converts the
	collection of $n_{\text{tasks}}$ models into a single model defined on the
	augmented input space. For a fixed task index $i$, the corresponding slice of
	$m$ coincides with the original model $m_i$.
	
	In practice, experimental and synthetic datasets with vector-valued outputs are
	mapped to the augmented input space by repeating each input once per task and
	associating it with the corresponding task index. The multivariate outputs are
	then stacked into a single scalar response vector, with one scalar observation per augmented input, yielding a representation
	compatible with the effective single-output model and a joint likelihood.
	
	For calibration cases that require GPs for the model, namely types $B$, $C$ and $D$, high-dimensional
	kernels need to be employed that do not introduce artificial coupling between different models. In
	other words, the covariance function in the extended space $\mathbb{R}^d\times[0,n_{tasks}-1]\times
	\mathbb{R}^p$ should vanish for pairs of data with different task indices. To illustrate this key idea,
	consider a type $B$ calibration. In a single-output calibration,  a covariance function
	$c_S:\mathbb{R}^d\times \mathbb{R}^d$ needs to be employed to define the model surrogate. In the
	multi-output case, a covariance $\hat{c}_S:\mathbb{R}^{d+1}\times \mathbb{R}^{d+1}$ is employed
	whose expression is simply
	\begin{equation}
		\label{eq-mo-covarianceB}
		\hat{c}_S(\hat{x},\hat{x}') =
		\begin{cases}
			c_S(x,x') & \textrm{if}\ i=j\ ,\\
			0         & \textrm{otherwise}\ ,
		\end{cases}
	\end{equation}
	where $\hat{x}=(x,i)$ and $\hat{x}'=(x',j)$ are the extended coordinates in $\mathbb{R}^{d+1}$.

	
	
	
	This block-diagonal construction results in a kernel that is active only within each task, treating
	outputs as independent and excluding cross-output correlations. Although computationally efficient,
	this design represents a strong assumption. In practice, tasks or outputs may share information, and
	further extensions could incorporate more expressive similarity kernels or downsampling kernels \cite{Wu2020,Mikkola2023,gantzler_multi-fidelity_2023,Ozdemiretal}, enabling correlations to be
	captured across outputs. It should be noted that in multi-output calibration,
	\emph{nondimensionalization and output scaling} are particularly critical. Since all outputs are
	treated with equal weight in this formulation, they must be normalized to comparable scales to
	ensure meaningful calibration results. 
	
	In this way, multi-objective calibration is included in \ACBICI. Accordingly, the four types of
	calibrations, with and without experimental error calibration, are extended to accommodate multiple,
	interrelated models depending on the same input variables and parameters. 
	
	\subsection{Numerical algorithms to approximate the posterior probability}
	\label{sec-posterior}
	All Bayesian inference calculations eventually require evaluating a posterior probability
	distribution. In calibration problems of the type investigated in this article, the computation
	of the posterior
	distribution of the model parameters is always the last step of the process. Due to the complexity
	of the likelihood functions (see Eqs.~\eqref{eq-typea-posterior}, \eqref{eq-posteriorB},
	\eqref{eq-typeC-bayes},  and \eqref{eq-typeD-bayes}), the posterior will (almost) never be
	analytically calculated and numerical methods will invariably be used. Here we review two of the
	most popular numerical methods for computing the posterior: Markov chain Monte Carlo (MCMC) sampling and Variational Bayesian Monte Carlo (VBMC).
	
	MCMC methods generate samples from the posterior distribution by constructing a Markov chain whose
	stationary distribution is the target posterior. After an initial burn-in phase, the chain yields
	approximate draws from the true posterior, enabling estimation of posterior moments, credible
	intervals, and derived quantities. MCMC makes minimal approximations and converges to the exact
	posterior in the limit of long chains. There is a vast literature on this extremely important
	algorithm that has been a cornerstone of applied mathematics for 75 years (see, for example,
	\cite{madras2002uh,robert2004gp,barbu2020is,romero2025qk}). Here, we refrain from describing the
	method yet we recall that MCMC works by evaluating the likelihood and the prior, keeping some
	samples and rejecting others. Even with the most recent and advanced variants of the method, and
	depending on the dimension of the sample space and the geometry of the posterior, thousands, tens of
	thousands, or more samples need to be evaluated. These evaluations entail a high computational cost.
	
	In the calibration problems addressed by \ACBICI, it is not uncommon to face inference problems that
	require a number of MCMC steps of the order of a hundred thousand. To exploit current multi-core
	processors, the library uses the \texttt{emcee} Python package \cite{Foreman_Mackey_2013}, an affine-invariant ensemble sampler that operates with multiple walkers. This approach is particularly effective for posteriors with correlated parameters or non-Gaussian geometry, as the ensemble proposal mechanism adapts automatically to the local covariance structure. However, even with the use of \texttt{emcee}, MCMC sampling can remain computationally expensive when likelihood evaluations are costly or when the posterior exhibits strong correlations or multimodality, potentially leading to slow mixing and long run times.
	
	Variational Bayesian Monte Carlo (VBMC) offers a sample-efficient and scalable alternative for posterior approximation, particularly when likelihood evaluations are computationally expensive. Following Acerbi~\cite{NEURIPS2020_5d409541}, VBMC models the unnormalized log-posterior
	\begin{equation}
		\label{eq-log-posterior}
		f(\Theta \vert \mathcal{D}) = \log L(\mathcal{D} \vert \Theta) + \log \pi(\Theta)
	\end{equation}
	with a GP surrogate, which provides uncertainty-aware predictions that guide exploration of the parameter space. 
	In Eq.~\eqref{eq-log-posterior} and below, $\Theta$ represents all the parameters and hyperparameters to be calibrated, which depend on the calibration type.
	
	In VBMC, the posterior~$f$ is approximated by a flexible variational family, typically a mixture of Gaussians of the form
	\begin{equation}
		q_{\phi}(\Theta) = \sum_{k=1}^{K} w_k\,\mathcal{N}(\Theta \mid \mu_k, \Sigma_k),
	\end{equation}
	whose parameters $\phi = \{w_k,\mu_k,\Sigma_k\}_{k=1}^K$ are optimized by maximizing an estimate of the evidence lower bound (ELBO),
	\begin{equation}
		\mathcal{L}(q_{\phi}) =
		\mathbb{E}_{q_{\phi}}\!\left[f(\Theta \vert \mathcal{D})\right]
		- \mathbb{E}_{q_{\phi}}\!\left[\log q_{\phi}(\Theta)\right],
	\end{equation}
	where expectations are computed using Monte Carlo samples and GP-based Bayesian quadrature. The GP surrogate is iteratively refined using an acquisition strategy inspired by Bayesian optimization, which selects informative parameter locations at which to evaluate the expensive likelihood. This combination of surrogate modeling, variational approximation, and active sampling enables VBMC to approximate both the posterior distribution and the model evidence with far fewer likelihood evaluations than traditional MCMC. Although VBMC introduces additional approximation through the surrogate model, it provides a powerful and computationally efficient inference method for complex models where direct MCMC sampling would be prohibitively expensive.
	
	The \ACBICI\ framework leverages the \texttt{PyVBMC} Python package to perform variational Bayesian Monte Carlo inference~\cite{Huggins2023,NEURIPS2020_5d409541,NEURIPS2018_747c1bcc}. Together, MCMC and VBMC form a robust inference toolkit for approximating the posterior probability distribution: MCMC enables accurate posterior characterization, whereas VBMC provides a scalable and sample-efficient approximation that is particularly advantageous when computational cost is a limiting factor.

	\subsection{Prediction with Gaussian process}\label{sec:pred}
	In types $B,C$ and $D$, GPs are used as surrogates for the model and/or the discrepancy. After the
	posterior probability distributions of the model parameters and the hyper-parameters of the GPs are
	calculated, it might be necessary to evaluate the GPs, to study uncertainty
	in the output, confidence intervals, maximum likelihood, etc. The evaluation of GPs demands a
	procedure that we have not covered before but, being at the core of the last step in model
	calibration, we summarize it next and refer to the literature for extended details~\cite[Algor.
	2.1]{rasmussen2006vz}. In contrast with the standard evaluation procedure for GPs, the one described
	next has some peculiarities that stem from the different nature of the experimental, synthetic, and
	prediction data. Specifically, only the experimental data might have measurement error and this is
	reflected, as shown below, in the structure of the joint covariance matrix.
	
	In all the calibration cases studied in this section, whenever a GP is used for the model or the
	discrepancy, a covariance function is introduced that gauges the correlation between a pair of input
	variables or a pair of input variables and parameters. This covariance function might include the
	effects of the experimental measurements. To present the evaluation of GPs of this type in a compact
	way let us write all the covariance functions employed in types $B,C,D$ as
	\begin{equation}
		\label{eq-eva-covariance}
		c(z_i,z_j) = \tilde{c}(z_i,z_j) + \sigma^2\, \delta^{exp}_{ij}\ .
	\end{equation}
	In this expression, each of the variables $z_i,z_j$ can represent an input, indicated before as~$x_i$, or
	a pair $(x_i,t_i)$ of input variables and model parameters. The experimental error only contributes to
	the covariance when $i=j$ and both of these points are obtained experimentally. Abusing slightly
	the notation, we use a Kronecker delta symbol to indicate this term.
	
	Let us assume that $z_i\in \mathcal{D}$, the set of $N$ points obtained either experimentally or
	synthetically for which the value $y$ of the model is known and has been used in the calibration. In
	addition, let us now calculate the value of the GP for new points in $\mathcal{P}=\left\{z^{*}_i,\ i=1,\ldots,P\right\}$. More
	precisely, let us calculate the expectation and variance of the GP at all $z^{*}_i\in\mathcal{P}$. For that, let us introduce the extended covariance matrix
	\begin{equation}
		\label{eq-extended-Sigma}
		\Sigma =
		\begin{bmatrix}
			\Sigma^{\mathcal{D} \mathcal{D}} & \Sigma^{\mathcal{D} \mathcal{P}} \\
			\Sigma^{\mathcal{P} \mathcal{D}} & \Sigma^{\mathcal{P} \mathcal{P}}
		\end{bmatrix} .
	\end{equation}
	The block $\Sigma^{\mathcal{D} \mathcal{D}}$ corresponds to the covariance between the data in
	$\mathcal{D}$ and has components
	\begin{equation}
		\label{eq-Sigma-DD}
		\Sigma^{\mathcal{D} \mathcal{D}}_{ij} = \tilde{c}(z_i,z_j) + \sigma^2\,\delta^{exp}_{ij}
		,
		\qquad\textrm{for}\ 1\le i,j\le N\ .
	\end{equation}
	The cross terms $\Sigma^{\mathcal{D} \mathcal{P}}= (\Sigma^{\mathcal{P} \mathcal{D}})^T$ account for
	the coupling between data and values to be predicted and is calculated as
	\begin{equation}
		\label{eq-Sigma-DP}
		\Sigma^{\mathcal{D} \mathcal{P}}_{ij} = \tilde{c}(z_i,z^{*}_j)
		,
		\qquad\textrm{for}\ 1\le i\le N,\ 1\le j\le P.
	\end{equation}
	Finally, the covariance block for the prediction is
	\begin{equation}
		\label{eq-Sigma-PP}
		\Sigma^{\mathcal{P} \mathcal{P}}_{ij} = \tilde{c}(z_i^{*},z_j^{*})\ ,
		\qquad
		\textrm{for}\ 1 \le i,j \le P.
	\end{equation}
	Collecting the $N$ outputs values of data $\mathcal{D}$ in the vector $y$, the predictions at 
	the inputs $z^{*}_i$ are gathered in the vector $y^{*}$ of length $P$ whose mean and covariance
	matrix are calculated as:
	\begin{equation}
		\begin{aligned}
			\mathbb{E}[y^*] &= \Sigma^{\mathcal{D} \mathcal{P}} [ \Sigma^{\mathcal{D} \mathcal{D}}]^{-1} y, \\
			\mathrm{Cov}[y^{*},y^{*}] &= \Sigma^{\mathcal{P} \mathcal{P}}-
			\Sigma^{\mathcal{P} \mathcal{D}}
			[\Sigma^{\mathcal{D} \mathcal{D}}]^{-1}
			\Sigma^{\mathcal{D}\mathcal{P}}.
		\end{aligned}
	\end{equation}

	
	\section{Practical recommendations}
	\label{sec-practical}
	When conducting practical Bayesian calibration studies, several factors can significantly influence
	both the efficiency and interpretability of the results. Attention to these details can enhance the
	reliability of inference and facilitate the broader applicability of the findings. We gather next
	some of the most important recommendations that are only found scattered through the literature and
	we point at how all of them are integrated in \ACBICI.
	
	\paragraph{Scaling}
	Isotropic kernels, such as the ones employed in Eqs.~\eqref{eq-covariance-kernel}
	and~\eqref{eq-rbf}, are functions of the distance between pairs of input variables or
	parameters. For these operations to be consistent, it is necessary that all the quantities in the
	distance function have the same dimensions.  Moreover, to enforce a similar sensitivity of the
	covariance with respect to all its inputs, it is beneficial to scale all of them so that their
	ranges are similar. A simple way to combine both goals is to non-dimensionalize and scale all input
	variables and parameters of the model prior to the calibration which, furthermore, has additional
	benefits from the numerical point of view, reducing the effects of floating point simplifications in
	the evaluation of exponential kernels. If input variables and parameters have dimensions, anisotropic
	kernels become mandatory \cite{menga2019ds,noack_autonomous_2020,hernandez-del-valle_robotically_2023}, introducing additional complexity associated with the
	estimation of the appropriate scaling dimension for each coordinate. In addition, sampling of the
	input space for the generation of synthetic data becomes cumbersome and the tuning of
	hyperparameters becomes more delicate.
	
	\paragraph{Selection of priors}
	Priors should reflect prior knowledge and stay within plausible bounds. Whenever reliable prior knowledge or expert judgment is available, it should be explicitly incorporated into the prior to improve inference and ensure physical or model-consistent results. Caution is required to ensure that the support of the prior includes all likely parameter values, otherwise certain values will be inadmissible. Empirical priors informed by data can be used but risk overfitting. The trade-off between uninformative and informative priors must be considered carefully, especially when synthetic or sparse data are involved -- uninformative priors risk leaving the posterior unchanged, while overly restrictive priors may rule out plausible solutions.
	
	\paragraph{Interpreting the plots}
	Diagnostic plots are central to evaluating the quality of Bayesian calibration results. 
	Trace, autocorrelation, and Gelman-Rubin statistic-/$\hat{R}$ plots provide insight into sampler
	mixing and convergence; the latter two are applicable only to MCMC methods and not to VBMC. Well-mixed chains, low autocorrelation times and values of $\hat{R}$ close
	to one, typically indicate successful MCMC sampling. Corner plots (also called pair plots) provide a comprehensive visualization of the joint and marginal posterior distributions of all parameters, revealing correlations, multimodality, and parameter identifiability. Prediction error plots visualize the distribution of errors in calibrated model predictions. These plots help identify systematic biases, heteroscedasticity, or outliers, providing insights into model adequacy and areas where the model may require refinement. The prior vs. posterior plots illustrate how much information the data have contributed relative to prior knowledge, which helps to assess identifiability and the influence of priors on the calibration results. Comparison plots that overlay experimental data with calibrated model predictions allow for a direct visual assessment of model fit quality across the input range, including the uncertainty quantified by Bayesian methods and surrogate-based predictions. Together, these diagnostic visualizations offer a multifaceted view crucial for reliable interpretation and validation of calibration studies.
	
	\paragraph{Convergence}
	Convergence diagnostics are essential to ensure that sampling algorithms provide reliable posterior
	estimates. Trace plots help assess whether Markov chains in the MCMC solution are well mixed and check for potential issues such as chains getting stuck or drifting. Autocorrelation plots, together with calculations of autocorrelation time, quantify the independence of samples in the chain. The integrated autocorrelation time quantifies the average number of iterations between effectively independent samples \cite{goodman_ensemble_2010}. This measure guides the estimation of the effective sample size (ESS). A high ESS indicates that the chain produces a large number of nearly independent samples, implying efficient exploration of the posterior and thus greater reliability of the estimates \cite{martino_effective_2017}. 
	
	The Gelman–Rubin statistic $\hat{R}$ \cite{gelman_inference_1992,rhat_revisiting_2021,Vehtari_2021} provides a complementary diagnostic for convergence,
	by comparing within-chain and between-chain variances. In \ACBICI, we use split-$\hat{R}$ 
	computations \cite{Vehtari_2021} that split each chain into two, doubling the number of chains analyzed. Values of
	$\hat{R}$ close to 1 suggest that chains have mixed well and converged to the same stationary
	distribution. To facilitate quick assessment, values are summarized in the library using a box plot, allowing the identification of parameters with $\hat{R}>1.01$ or $\hat{R}>1.05$, which may indicate poor convergence and require longer sampling or reparameterization.
	
	For variational inference with VBMC, convergence is assessed in \ACBICI\ using diagnostics tailored
	to the optimization-based posterior approximation rather than MCMC chain behavior. During the
	iterative refinement of the variational posterior, VBMC tracks the evidence lower bound (ELBO) and
	its estimated uncertainty; convergence is indicated when the ELBO plateaus and changes during
	iterations become small relative to its standard deviation. In parallel, VBMC monitors the symmetrized KL divergence between successive variational posteriors, which should decrease toward zero as the approximation stabilizes. These criteria are combined into a reliability (convergence) index $r_{index}$. In \texttt{PyVBMC}, inference is terminated only once $r_{index}<1$ and remains below this threshold for a prescribed stable-count window, indicating that both the objective and the variational posterior have become stable.
	
	In practice, low autocorrelation times, high ESS, and $\hat{R}$ values close to 1 collectively indicate efficient sampling and reliable posterior inference for MCMC, or \texttt{emcee}, respectively, while a stabilized ELBO, near-zero inter-iteration sKL, and a low reliability index provide analogous evidence of robust convergence for VBMC. In well-behaved analyses, marginal posterior distributions should exhibit clear peaks for most quantities, signalling that the algorithms have adequately explored (MCMC, \texttt{emcee}) or accurately approximated (VBMC) the main modes of the parameter space. Employing these diagnostics collectively gives confidence that the calibration has converged and the results are robust and reflective of the true posterior distribution.
	
	\paragraph{Sensitivity}
	If sampler convergence or efficiency is problematic, one should consider narrowing parameter bounds,
	refining priors, or fixing parameters with low impact. This strategy focuses the calibration on the most influential quantities and can improve both performance and interpretability.
	
	\section{Examples}
	\label{sec-examples}
	In this section, we present practical calibration examples that cover the main problem types introduced in Section~\ref{sec-background}. Each example specifies the model and calibration setting, outlines how it is implemented with \ACBICI, and summarizes the resulting inferences and diagnostics. We emphasize the connection between the theory of Bayesian calibration and the library elements, to facilitate adaptation to other applications. All data and complete example scripts are included in the \ACBICI\ distribution and documented in the online manual.

	\subsection{Estimating gravity}
	\label{subs-gravity}
	This example illustrates the calibration of a simple physical model that has been 
	previously studied in the literature~\cite{mcclarren2018yq}. Specifically, the example shows how
	Bayesian calibration can be employed to infer the probability distribution of a physical constant,
	gravity in this case, from prior information, experimental measurements, and a model. Two
	calibration types are illustrated here: a case $A$ calibration, as in Section~\ref{subs-caseA}, the simplest of all, and a case $C$
	calibration, as in Section~\ref{subs-casec}, where the discrepancy is calculated.
	
	In the example, $N$ drop tests are used to estimate the gravitational acceleration $g$. For that, an
	object is dropped from heights $h_i,\ i=1,\ldots,N$ and the times it takes to the floor, indicated as
	$t_i$, are measured and collected in the data set $\mathcal{D}=\{(h_i,t_i), \ i=1,\ldots, N\}$. The
	gravitational acceleration $g$ is the physical constant that, if assumed constant, relates $h$ and
	$t$ through the model
	\begin{equation}
		\label{eq-gravity-t}
		t = \sqrt{\frac{2h}{g}}.
	\end{equation}
	
	This example has been selected to illustrate the features of the framework, and the drop test data are not obtained from lab tests but generated, instead, with a program that allows us to vary the noise in the data and the deviation of the ``measurements'' from the predictions~\eqref{eq-gravity-t}. To push the features of the Bayesian framework, the data is generated using a model different from Eq.~\eqref{eq-gravity-t}.

	To calibrate this model with \ACBICI, it is required to collect the experiments,
	define the model, and select a prior probability distribution for the only unknown parameter,
	namely~$g$. In this example, data are taken from the file \texttt{experiments.dat}, distributed with the library, and the model is defined as a subclass of the \texttt{ACBICImodel} parent class. In the constructor, the dimension of
	the input space ($d=1$), the name of the parameter to be estimated ($g$), and its prior (in this case a uniform distribution over $[7,12]$) are given. By default, the calibration analysis infers the standard
	deviation $\sigma$ of the experimental error. Then, the analytical expression of the
	model~\eqref{eq-gravity-t} is implemented as:
	\begin{verbatim}
		class myModel(ACBICImodel):
		def __init__(self):
		self.xdim = 1
		self.addParameter(label=r'$g$', 
		prior=Uniform(a=7.0, b=12.0))
		
		def symbolicModel(self, x, p):
		x = to_column_vector(x)
		p = to_column_vector(p)
		r = np.sqrt(2.0*x[:,0]/p[:,0])
		r = to_column_vector(r)
		return r
	\end{verbatim}

	\subsubsection{Classical calibration with MCMC sampling}
	\label{ss-a-mcmc}
	In \ACBICI, type~A calibration is performed using the \texttt{classicalCalibrator} class. In this simple inference problem, the library just needs to be fed with the experimental data and linked to the symbolic model defined before. This is done with the commands:
	\begin{verbatim}
		experiments = np.loadtxt("data/experiments.dat")
		theCalibrator = classicalCalibrator(theModel)
		theCalibrator.storeExperimentalData(experiments)
		theCalibrator.calibrate()
		theCalibrator.plot()
	\end{verbatim}
	This short program produces a calibration report with standard MCMC diagnostics (effective sample sizes, split-$\hat{R}$, autocorrelation times), posterior summaries, and prediction error statistics, as well as a set of plots.
	\begin{figure}[t]
		\centering
		\includegraphics[width=.85\linewidth]{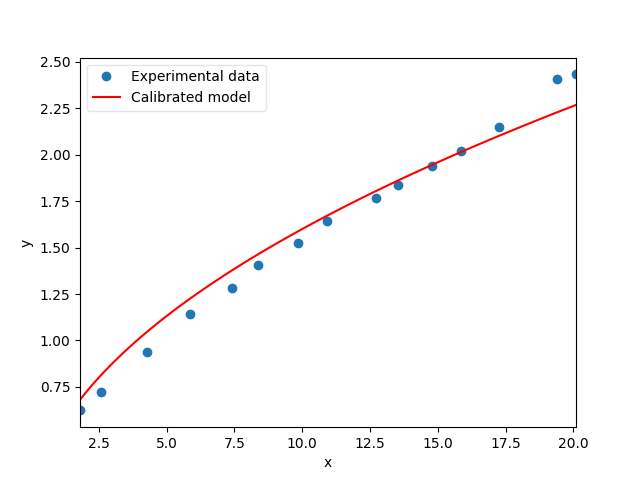}
		\caption{Example~\ref{ss-a-mcmc}. Comparison between experimental data (blue markers) and calibrated model predictions (red line). Close agreement indicates accurate reproduction of the observed behavior using the MAP parameter estimates.}
		\label{fig:comparecornerA0a}
	\end{figure}
	\begin{figure}[htb]
		\centering
		\includegraphics[width=.85\linewidth]{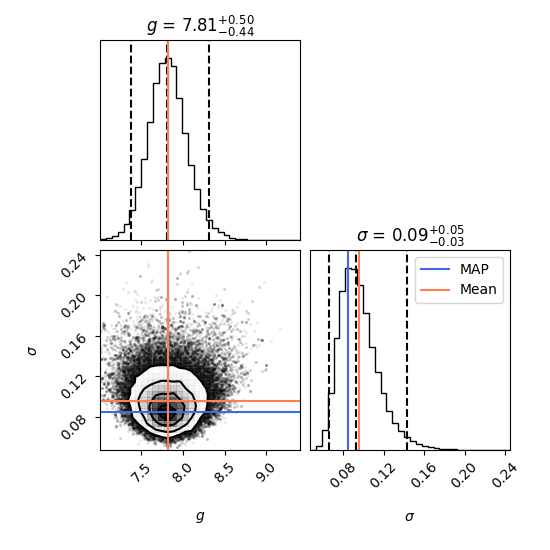}
		\caption{Example~\ref{ss-a-mcmc}. Corner plot showing marginal and joint posterior distributions of the calibrated parameters. Diagonal panels display marginal densities with MAP, mean, median, and 95\% credible intervals; off-diagonal panels illustrate parameter correlations.}
		\label{fig:comparecornerA0b}
	\end{figure}
	One of the obtained plots is shown in Figure~\ref{fig:comparecornerA0a}. It compares the experimental data with the calibrated model predictions obtained from the maximum a posteriori (MAP) estimate of $g$, showing that the model accurately reproduces the observed fall times. Figure~\ref{fig:comparecornerA0b} displays the marginal and joint posterior distributions for the calibrated parameters, confirming that $g$ is well-identified and that the posterior is substantially more concentrated than the uniform prior specified in this example. The MCMC diagnostics indicate good chain mixing and convergence; detailed numerical values and additional plots are reported in the Supplementary Information.
	
	The experimental data for this example are provided in SI units. Based on them, the posterior probability distribution of gravity obtained by \ACBICI\ shows a MAP close to $7.8$ \si{m/s^2}. This unexpectedly small value for gravity is the mode of the distribution of $g$ that best explains the ``experimental'' measurements, given its prior distribution, \emph{as long as model~\eqref{eq-gravity-t} is an accurate predictor of falling times}. 
	
	In this simple example, we know that the MAP is a bad estimate of the real value of gravity. In fact, the true reason for this apparently poor result is the inadequacy of the model~\eqref{eq-gravity-t} to measure falling times of certain objects, as we will discuss next. We stress that, however, in general we do not have the luxury of knowing the true value of the parameter so type $A$ calibration might be deceiving. The conclusion of this analysis is that unless we are sure that the model we employ is an accurate predictor of the experimental data, we should use more sophisticated calibration methods that can identify the model discrepancy.
	
	\subsubsection{Discrepancy calibration with VBMC}
	\label{ss-c-bayes}
	In type~$A$ calibration, the probability distribution of the unknown parameter is selected to explain, as well as possible, the experimental data obtained from the drop tests. However, the model ignores effects such as air friction, and the expression~\eqref{eq-gravity-t} might not be appropriate for very light objects that are not aerodynamic, such as leaves, soap bubbles, etc. For these and similar cases, the best parameter distribution might not be good enough, as we have seen.
	
	We now account for possible structural inadequacies in the model by introducing a discrepancy term, leading to a type~$C$ calibration. In this analysis, there is no attempt to find the functional form of
	the discrepancy, but merely separate the latter from the calibration of the sought parameter, and therefore improve the calibration. 
	
	The discrepancy is modeled with a Gaussian process surrogate and incorporated into the
	calibration using a \texttt{discrepancyCalibrator} object. To accelerate the computations, we 
	select the variational Bayesian Monte Carlo (VBMC) to estimate the posterior using:
	\begin{verbatim}
		experiments = np.loadtxt("data/experiments.dat")
		theCalibrator = discrepancyCalibrator(theModel, name="typeC")
		theCalibrator.storeExperimentalData(experiments)
		theCalibrator.calibrate(method="vbmc")
		theCalibrator.plot()
	\end{verbatim}
	
	Figure~\ref{fig:comparecornerC0a} shows the predictions of the calibrated model alongside the 
	data employed for the calibration of its parameters. The calibrated model~\eqref{eq-gravity-t} enriched with discrepancy predicts the data points with accuracy. However, if the discrepancy corrections are removed, the predictions of the calibrated model are worse. This deterioration of the predictions should be expected: the Gaussian process that models the discrepancy absorbs the systematic model/data mismatch as seen in Figure~\ref{fig:discrepancyCplot}. Figure~\ref{fig:comparecornerC0b} shows the pairwise and marginal probability distributions for the parameter $g$, the experimental error $\sigma$, and remaining hyperparameters. When compared with the probability distributions of Figure~\ref{fig:comparecornerA0b}, we notice that the experimental error in type $C$ is now estimated to be smaller, since the cause for mismatch is attributed now to model discrepancy, not to measurement errors. As a direct consequence, the probability distribution of $g$ becomes broader: the uncertainty in the model precludes a well-defined value for the parameter. This apparent loss of information on the parameter distribution is actually an advantage of the approach: the method predicts a broader interval of confidence for gravity, one that includes the real value, because it identifies that the experiments do not fit well with the predictions of the simple model~\eqref{eq-gravity-t}. Other information, such as VBMC diagnostics (ELBO evolution, KL divergence, convergence index) indicate a stable variational solution; full numerical summaries and diagnostics are provided in the Supplementary Information.
	
	\begin{figure}[ht]
		\centering
		\includegraphics[width=0.85\linewidth]{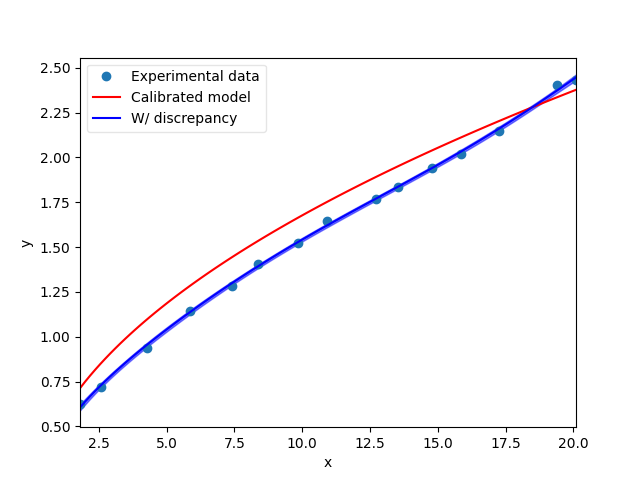}
		\caption{Example~\ref{ss-c-bayes}. Comparison between experimental data (markers) and model predictions after discrepancy calibration. The discrepancy-corrected predictions provide improved agreement with the observations, particularly in regions where the base model exhibits systematic deviations.}
		\label{fig:comparecornerC0a}
	\end{figure}
	\begin{figure}[ht]
		\centering
		\includegraphics[width=0.85\linewidth]{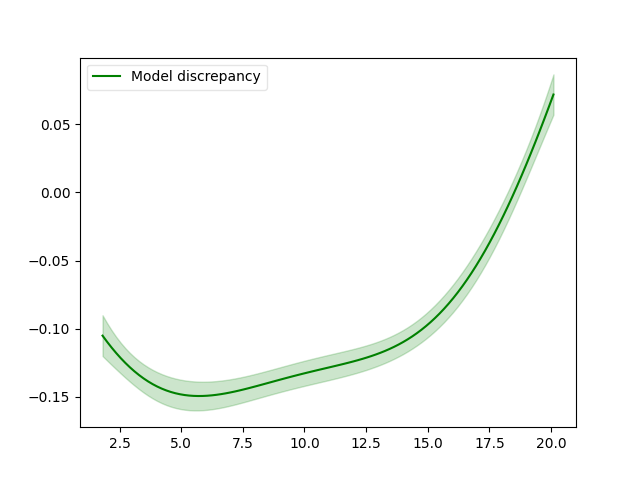}
		\caption{Example~\ref{ss-c-bayes}. Estimated model discrepancy using a Gaussian process surrogate, reflecting systematic deviations between the model and data.}
		\label{fig:discrepancyCplot}
	\end{figure}
	\begin{figure}[htb]
		\centering
		\includegraphics[width=0.85\linewidth]{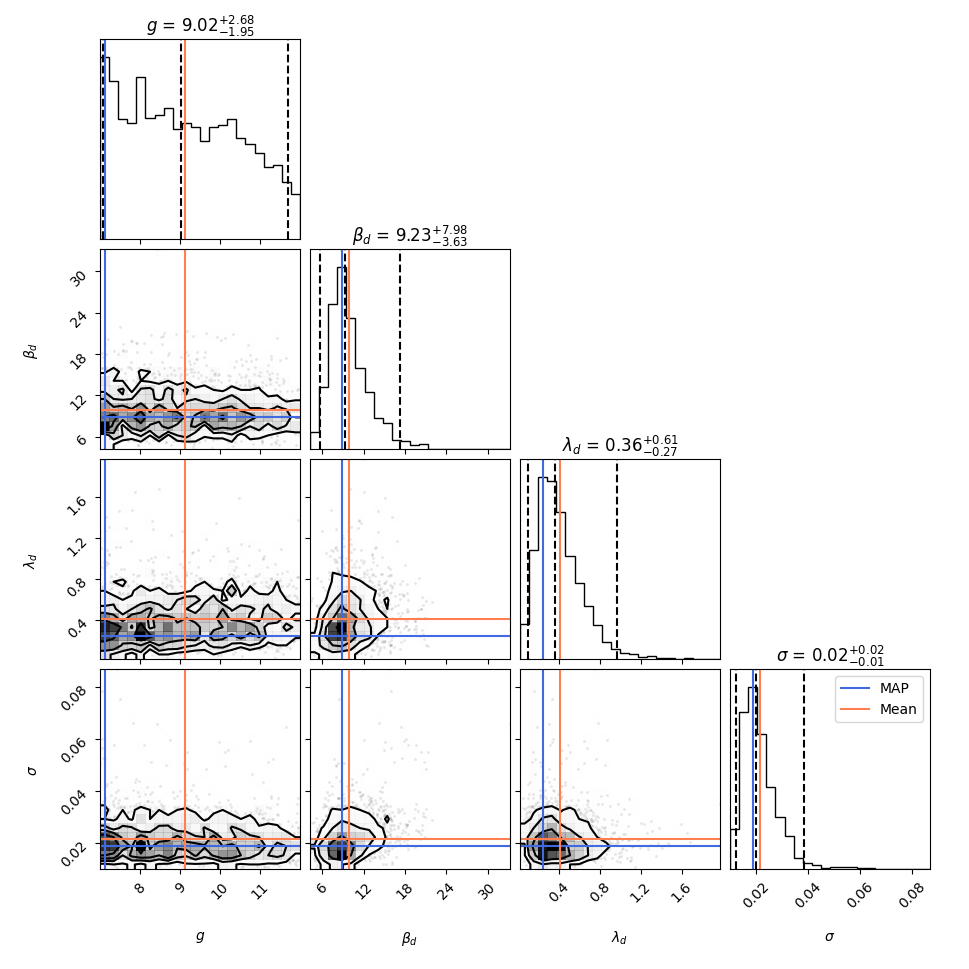}
		\caption{Example~\ref{ss-c-bayes}. Posterior distributions and parameter correlations obtained via variational Bayesian Monte Carlo. The widened distribution for \(g\) reflects the uncertainty introduced by explicitly modeling structural discrepancy, while the additional hyperparameters characterize the inferred discrepancy process.}
		\label{fig:comparecornerC0b}
	\end{figure}

	\subsection{Cobb–Douglas production model calibration}
	\label{subs-cobb}
	We consider now an example brought from Economics. We study the Cobb–Douglas production function \cite{cobb_theory_1928,barro_economic_2004} that relates total output $Y$ of a certain good to inputs of labor $L>0$, capital spent $K>0$, and a productivity factor $T\in[0,1]$:
	\begin{equation}
		\label{eq-cobb-douglas}
		Y = T \, L^{\alpha} K^{\gamma},
	\end{equation}
	where $\alpha$ and $\gamma$ are the output elasticities with respect to labor and capital, two parameters that need to be calibrated based on collected measurements.
	Here, $Y, L$, and $K$ are measured in economic units (e.g. monetary units or normalized indices), $T$ is dimensionless, and $\alpha, \gamma$ are dimensionless output elasticities.
	
	As suggested in Section~\ref{sec-practical}, numerical stability of the calibration procedure is improved when the input variables are non-dimensionalized.
	To effect this scaling, data for $L$ and $K$ are divided, respectively, by two reference values $L_0 = 50$ and $K_0 = 200$, that economic unit and are obtained from the available experimental data. Then, we define the scaled variables
	\begin{align}
		\hat{L} &= \frac{L}{L_0}, & \hat{K} &= \frac{K}{K_0}.
	\end{align}
	The productivity factor $T$ is already dimensionless so there is no need to scale it. From this point, we perform the calibration of the model using the scaled input vector $x = (T, \hat{L}, \hat{K})$.
	
	\subsubsection{Model definition}
	The first step in every calibration is to describe the model that needs to be calibrated and the parameters upon which it depends. In \ACBICI, this task is performed by deriving from the \texttt{ACBICImodel} class a new one, indicating the dimension of the input space and declaring each parameter to be calibrated, together with its attendant prior probability distribution:
	\begin{verbatim}
		class myModel(ACBICImodel):
		def __init__(self):
		self.xdim = 3
		self.addParameter(label=r'$\alpha$', 
		prior=Uniform(a=0.3, b=1.6))
		self.addParameter(label=r'$\gamma$', 
		prior=Normal(mu=0.35, sigma=0.1))
		
		def symbolicModel(self, x, p):
		T = x[:,0]; L = x[:,1]; K = x[:,2]
		alpha = p[:,0]; gamma = p[:,1]
		r = T * L**alpha * K**gamma
		return r.reshape(-1, 1)
	\end{verbatim}
	
	\subsubsection{Calibration procedure}
	For the calibration of the Cobb-Douglas model, we will assume that it is computationally expensive. Thus, we will follow a type $B$ procedure (see Section~\ref{subs-caseb}) that replaces the actual model with an inexpensive GP surrogate. In addition, we will like to find the variance of the experimental error, whose prior is assumed to be a random variable with a Gamma probability distribution. The \ACBICI\ code that takes care of all these steps is simply:
	\begin{verbatim}
		experiments = np.loadtxt("data/experiments.dat")
		calibratorBexp = expensiveCalibrator(theModel, name="calBE")
		calibratorBexp.setExperimentalSTDPrior(Gamma(alpha=2, beta=2))
		calibratorBexp.storeExperimentalData(experiments)
		calibratorBexp.calibrate(nsteps=15000)
	\end{verbatim}
	The full script and configuration details are provided in the Supplementary Information and \texttt{ACBICI}'s distribution. 
	
	Note that by replacing the true model~\eqref{eq-cobb-douglas} with a surrogate, we unavoidably introduce uncertainty in the calibration. As a result, the posterior distributions of the calibration parameters will have a larger spread than if we have employed simply a type $A$ methodology. In general, one must weigh the advantages of performing a (relatively) fast calibration against the drawback of obtaining a less sharp calibration. In the most extreme case, it might happen that the number of model evaluations available is fixed and a type $B$ calibration (or its extension, type $D$) becomes mandatory.
	
	\begin{figure}[htb]
		\centering
		\includegraphics[width=0.85\linewidth]{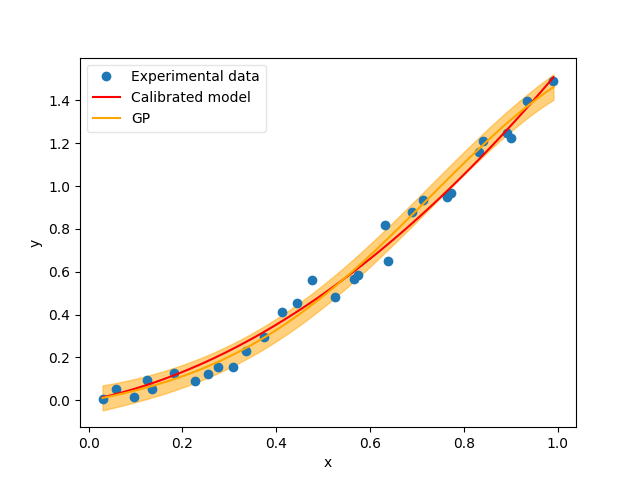}
		\caption{Example~\ref{subs-cobb}. Comparison of experimental data for the first input dimension (markers), calibrated model predictions (red), and uncertainty band for the GP-based surrogate (orange). The model and surrogate closely follow the observed trend, indicating successful calibration and accurate surrogate representation.}
		\label{fig:compare3a}
	\end{figure}
	
	\begin{figure}[htb]
		\centering
		\includegraphics[width=0.99\linewidth]{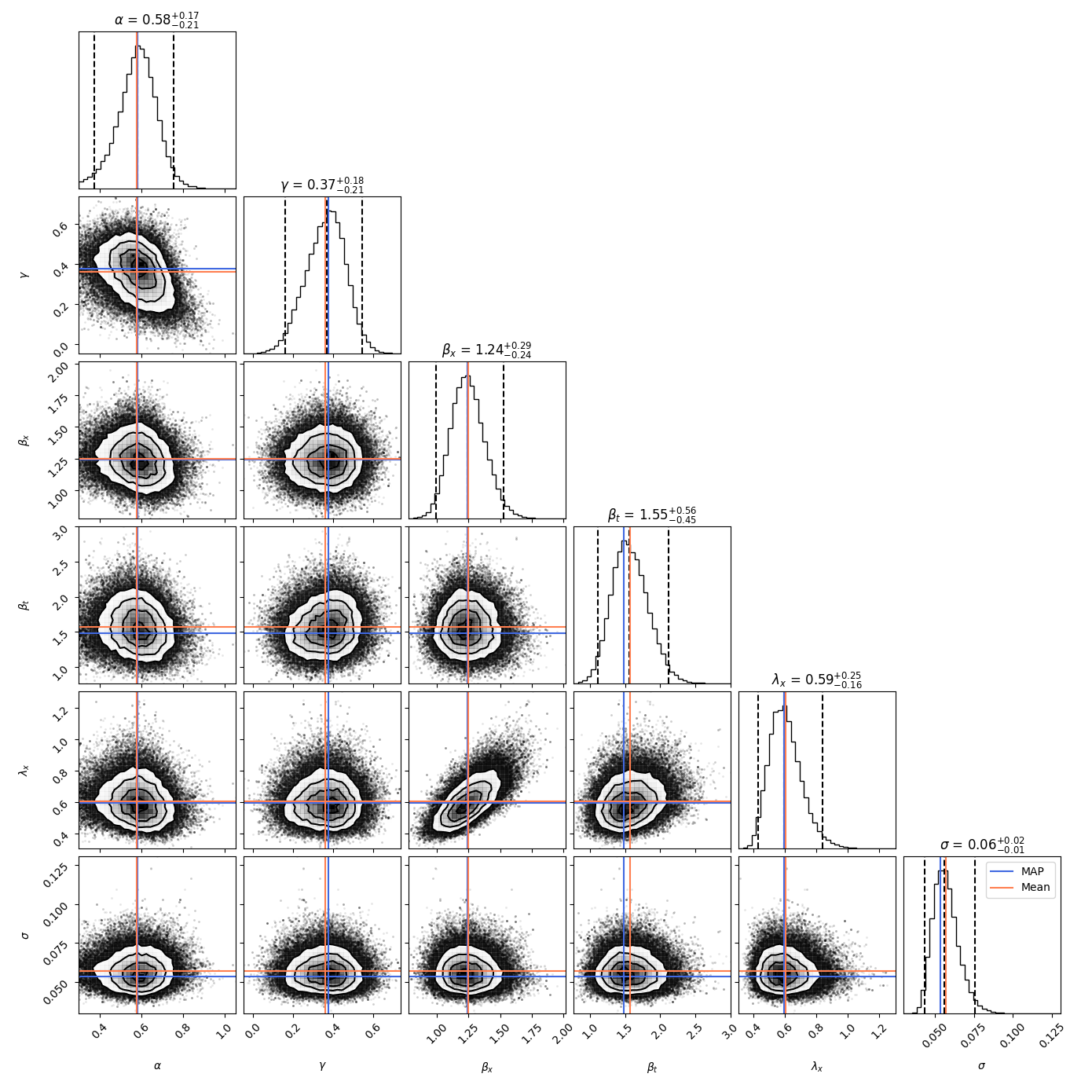}
		\caption{Example~\ref{subs-cobb}. Posterior distributions and pairwise correlations for the calibrated parameters \(\alpha\), \(\gamma\), and the GP hyperparameters. Marginal distributions show credible intervals and parameter uncertainty, while joint distributions reveal dependencies.}
		\label{fig:corner3b}
	\end{figure}
	Figures~\ref{fig:compare3a} and \ref{fig:corner3b} show that the calibrated model and surrogate accurately reproduce the data, with well-identified elasticities that have means $\alpha \approx 0.58$ and $\gamma \approx 0.36$, respectively, and tight posteriors for GP hyperparameters. MCMC diagnostics confirm good convergence; detailed numerical summaries are in the Supplementary Information. 
	
	Figure~\ref{fig:compare3a} shows the mean function of the surrogate GP, evaluated with the MAP of the calibrated parameters and hyperparameters, as well as a colored region that shows the probability is concentrated, for each value of the first input dimension. Also, the value of the original model is shown with its parameters given the MAP values of the calibration. Remarkably, the calibrated model --- not having been used for the actual calibration but only for setting up the meta-model --- closely approximates the data, pointing to the fact that the calibration has been successful.

	\subsection{Multi-objective traction test calibration}
	\label{subs-multiobjective}
	\begin{figure}[htb]
		\centering
		\resizebox{0.8\textwidth}{!}{
			\begin{tikzpicture}[line join=round, line cap=round, >=Latex]
				
				\def\L{6.0}     
				\def\H{1.3}     
				\def\D{1.0}     
				\def\shift{0.9} 
				
				\coordinate (FBL) at (\L,0);          
				\coordinate (FBR) at (\L+\D,0.4);     
				\coordinate (FTR) at (\L+\D,\H+0.4);  
				\coordinate (FTL) at (\L,\H);         
				
				\coordinate (BBL) at (0,0);
				\coordinate (BBR) at (\D,0.4);
				\coordinate (BTR) at (\D,\H+0.4);
				\coordinate (BTL) at (0,\H);
				
				\pgfmathsetmacro{\Hwall}{\H+0.8}
				\pgfmathsetmacro{\Hhatch}{\H+0.6}
				
				\draw[thick] (0.0,0.3) -- (0.0,\Hwall);
				
				\foreach \y in {-0.2,0.1,...,\Hhatch}{
					\pgfmathsetmacro{\yshift}{\y+0.4}
					\draw[thick] (-.3,\y) -- (0.,\yshift);
				}
				
				\draw[thick] (BBL) -- (FBL) -- (FTL) -- (BTL) -- cycle;  
				
				\draw[thick] (FBR) -- (FTR) -- (BTR);
				
				\draw[thick] (FBL) -- (FBR);  
				\draw[thick] (FTL) -- (FTR);  
				\draw[thick] (BTL) -- (BTR);  
				
				\fill[gray!8] (FBL) -- (FBR) -- (FTR) -- (FTL) -- cycle;
				\node[font=\large] at ($(FBL)!0.38!(FTR)$) {$A$};
				
				\draw[very thick, -{Latex[length=4mm]}]
				($(FBL)!0.5!(FTL) + (0.5,0.2)$) -- ++(3.0,0.0)
				node[right, font=\large] {$F$};
				
				\tikzset{hidden/.style={thick,dashed,gray!70}}
				
				\draw[hidden] (BBL) -- (BBR);
				\draw[hidden] (BBR) -- (BTR);
				\draw[hidden] (BBR) -- (FBR);
				\draw[thick] (0,-0.9) -- (0,-0.2);
				\draw[thick] (\L,-0.9) -- (\L,-0.2);
				
				\draw[thick, <->]
				(0,-0.75) -- (\L,-0.75)
				node[midway, fill=white, inner sep=1pt, font=\large] {$L$};
				
		\end{tikzpicture}}
		\caption{Example~\ref{subs-multiobjective}. Traction test schematic.}
		\label{fig:ex4sketch}
	\end{figure}
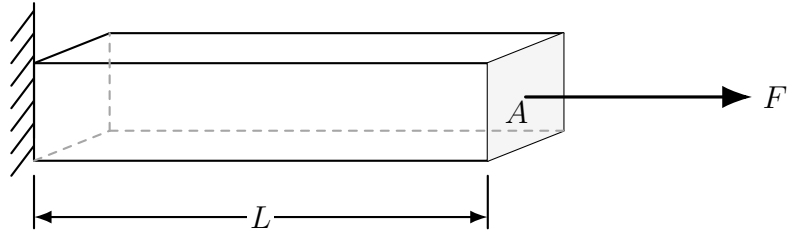
	
	In this last example, we explore the calibration of a multi-objective model. Specifically, we consider the common problem faced in Mechanics where Young's modulus and Poisson's ratio of a certain elastic material have to be inferred from a traction test  (see Figure~\ref{fig:ex4sketch}). Specifically, a prismatic bar of length $L_0=10$ \si{mm} and cross section area $A_0=1$ \si{mm^2} is pulled with a force~$F$. According to the theory of elasticity, the length and cross section area after the application of the force have, respectively, the expression
	\begin{equation}
		\label{eq-traction-test}
		y_1 = L \left( 1 + \frac{F}{A E} \right), \quad
		y_2 = A \left( 1 - \frac{F \nu}{A E} \right)^2,
	\end{equation}
	with $E,\nu$ being the  Young's modulus and Poisson's ratio of the material, respectively. The outputs $y_1$ and $y_2$ have units of length and area, respectively; $E$ has units of pressure, while $\nu$ is dimensionless. Note that the two models in Eq.~\eqref{eq-traction-test} depend both on the same input variable $F\in\mathbb{R}$ and parameter $E\in \mathbb{R}^+, \nu\in[0,1/2)$.
	
	Since the outputs $y_1,y_2$ have different dimensions, it proves mandatory to non-dimensionalize the equations, before any calibration. For that, we define the nondimensional length $\hat{L}=L/L_0$, area $\hat{A}=A/A_0$ and stiffness $\hat{E}=E/E_0$ with $E_0=250$ \si{MPa}. Then, Eq.~\eqref{eq-traction-test} can be written in non-dimensional form as
	\begin{equation}
		\hat{y}_1 = \hat{L} \left(1 + \frac{\hat{F}}{\hat{A} \hat{E}}\right), \quad
		\hat{y}_2 = \hat{A} \left(1 - \frac{\hat{F} \nu}{\hat{A} \hat{E}}\right)^2,
	\end{equation}
	with $\hat{y}_1=y_1/L_0,\ \hat{y}_2=y_2/L_0$ and $\hat{F}/(E_0\,A_0)$. Note that, after this scaling, the calibrated values of the two unknown model parameters, namely $\hat{E}$ and $\nu$, should be of the order of~1.
	
	\subsubsection{Model definition}
	In \ACBICI, the calibration of multi-output problems follows a similar structure than in single-output cases. The initial step, irrespective of the calibration strategy later selected, is to define the model to be calibrated, including the dimensions of the input and output spaces. Then, as in the examples of Sections~\ref{subs-gravity} and~\ref{subs-cobb}, the parameter names and prior probability distribution must be defined:
	\begin{verbatim}
		class myModel_nondim(ACBICImodel):
		def __init__(self):
		self.xdim = 1
		self.ydim = 2
		E0 = 250.
		self.addParameter(label=r'$E$', 
		prior=Uniform(a=200.0/E0, b=400.0/E0))
		self.addParameter(label=r'$\nu$', 
		prior=Uniform(a=0.0, b=0.5))
		
		def symbolicModel(self, x, p):
		r1 = L_hat * (1 + x[:,0] / (A * p[:,0]))
		r2 = A_hat * (1 - x[:,0] * p[:,1] / (A * p[:,0]))**2
		return np.column_stack([r1, r2])
	\end{verbatim}

	\subsubsection{Calibration procedure}
	After the model is defined, any calibration type can be selected for the multi-output calibration. Here, we have decided to use a type $D$ method that models the discrepancy and employs a surrogate of the symbolic model. In addition, we assume that there are experimental errors in the measurements employed and we aim to estimate the variance of their distribution, assumed as in all other examples to be Gaussian and of zero mean. In \ACBICI, type~$D$ calibration is implemented through the \texttt{KOHCalibrator} that loads the experimental data available and then carries out the whole procedure with five commands:
	\begin{verbatim}
		experiments = np.loadtxt("data/experiments.dat")
		theCalibrator = KOHCalibrator(theModel, name="calDE")
		theCalibrator.storeExperimentalData(experiments)
		theCalibrator.storeSyntheticData(sd)
		theCalibrator.calibrate(nsteps=20000)
	\end{verbatim}
	The full script is in the Supplementary Information.
	\begin{figure}[t]
		\centering
		\includegraphics[width=0.85\linewidth]{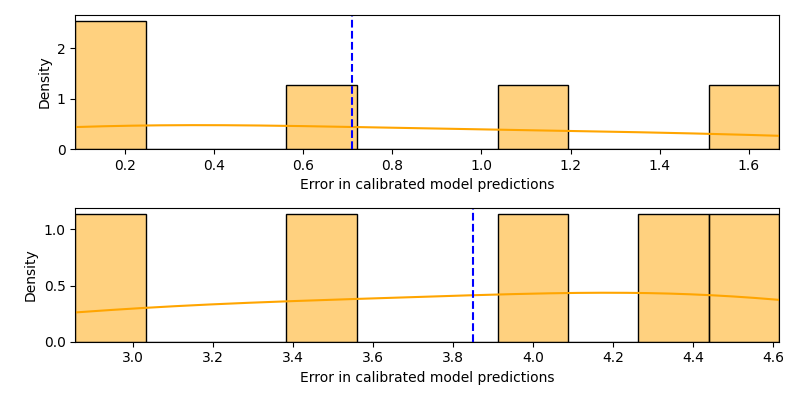}
		\caption{Example~\ref{subs-multiobjective}. Prediction error plot showing residuals for both outputs across the input range. Clustering of residuals near zero indicates good predictive performance, while broader spreads highlight regions of increased uncertainty or model mismatch.}
		\label{fig:predictions4a}
	\end{figure}
	
	\begin{figure}[t]
		\centering
		\includegraphics[width=0.99\linewidth]{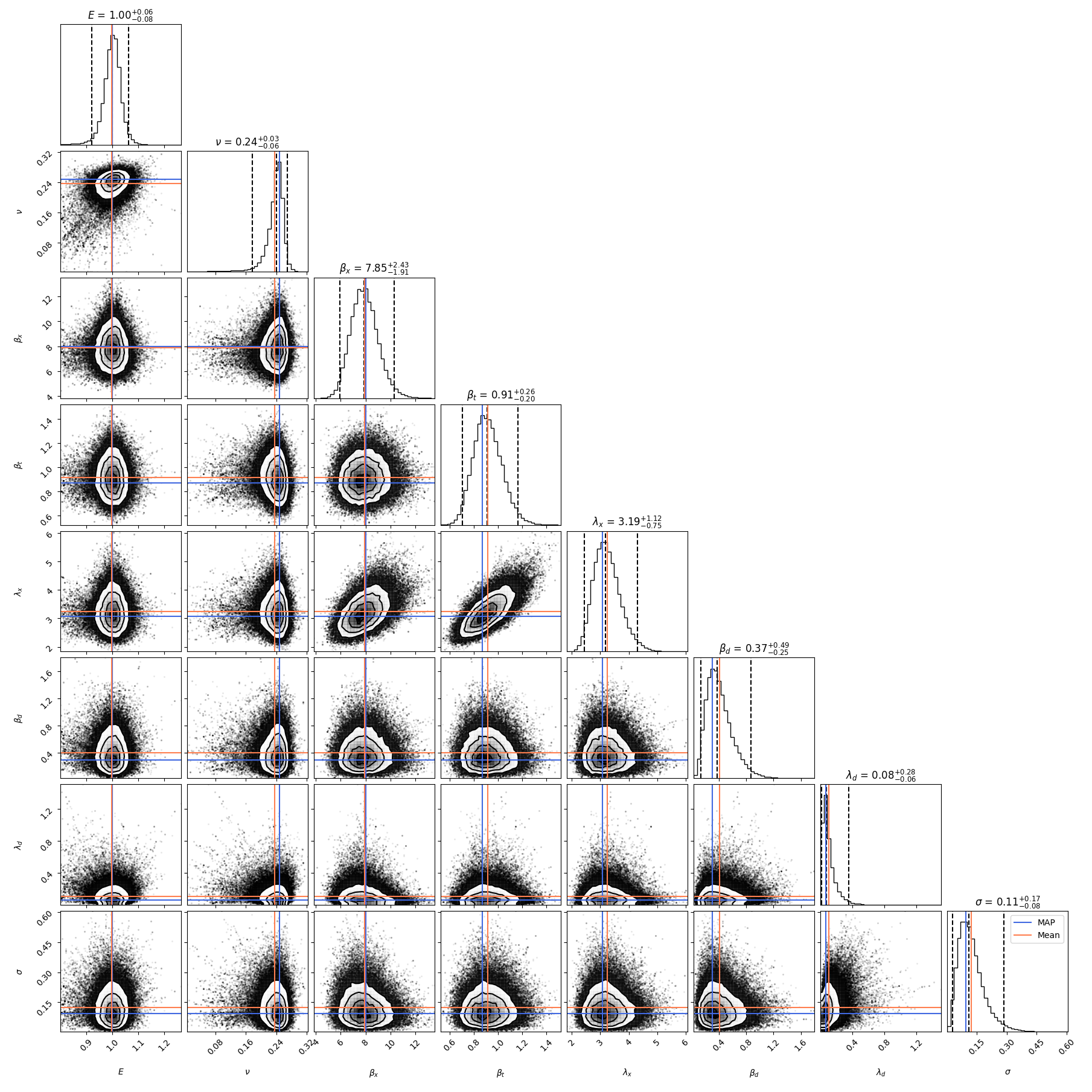}
		\caption{Example~\ref{subs-multiobjective}. Posterior distributions and pairwise correlations for all calibrated and surrogate-related parameters. Marginal distributions provide credible intervals and uncertainty quantification, while joint distributions reveal parameter dependencies and identifiability.}
		\label{fig:corner4b}
	\end{figure}
	
	\begin{figure}[t]
		\centering
		\includegraphics[width=0.85\linewidth]{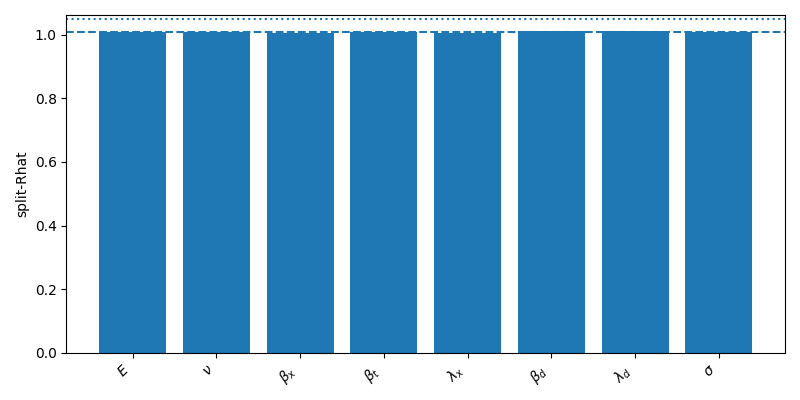}
		\caption{Example~\ref{subs-multiobjective}. Split-\(\hat{R}\) convergence diagnostics for all calibrated and surrogate-related parameters.
			Bars correspond to individual parameters (\(E\), \(\nu\), \(\beta_x\), \(\beta_t\), \(\lambda_x\), \(\beta_d\), \(\lambda_d\), and \(\sigma\)),
			while the two dashed lines mark the recommended convergence thresholds of $1.01$ and $1.05$.
			Values very close to 1 indicate that all chains have mixed well and that no parameter shows signs of non-convergence.}
		\label{fig:rhat4}
	\end{figure}
	
	Figures~\ref{fig:predictions4a}–\ref{fig:rhat4} show residuals clustered near zero across outputs, well-constrained posteriors for \(E \approx 1\) and \(\nu \approx 0.25\), and split-$\hat{R}$ values near 1 for all parameters, respectively, confirming good multi-output fit and MCMC convergence. Detailed information and diagnostics are in the Supplementary Information.
	
	The examples in this section do not attempt to investigate the corresponding three models. Rather, they intend to show that all the potential calibration strategies reviewed in Section~\ref{sec-background} can be tackled in a straightforward and unified fashion with \ACBICI. Irrespective of whether the model to calibrate is expensive or inexpensive, adequate or not to represent the available data, with one or multiple outputs and with known or unknown experimental error, calibrating it amounts to describing the model and then employing a short list of Python commands provided by the library.

	\section{Conclusions and main results}
	\label{sec-conclusions}
	In this work, we present a detailed review of the theoretical foundations of Bayesian calibration in the Kennedy–O’Hagan (KOH) framework, with a particular focus on uncertainty quantification, model discrepancy, and posterior inference. Building on this exposition, we introduce \ACBICI~\cite{ACBICIcode,ACBICIdocs}, a Python package designed to mirror the key concepts and methods underlying KOH calibration while providing a practical and extensible computational implementation.
	
	The framework supports both single- and multi-output calibration, extending the classical KOH formalism to enable simultaneous Bayesian inference across multiple, interrelated model outputs. Advanced Bayesian inference methods, posterior approximation techniques, Gaussian process–based prediction, and sensitivity analysis are integrated into a unified and automated workflow, facilitating scalable, interpretable, and sample-efficient inference for complex models, including in data-limited settings.
	
	In addition, we assemble a set of practical recommendations that translate theoretical considerations and implementation experience into actionable guidance for conducting reliable Bayesian calibration without sacrificing methodological rigor.
	
	Through a series of illustrative examples from applied science and engineering, we demonstrate robust uncertainty quantification, improved model interpretability, and practical applicability to real-world scenarios. The combination of advanced inference techniques with intuitive workflow automation and ease of use lowers the barrier to adopting sophisticated calibration procedures, while sensitivity analysis tools assist users in identifying influential parameters and guiding efficient model refinement.
	
	Overall, this work bridges statistical theory and practical computational modeling, offering a structured reference on KOH methodology together with practical recommendations and a user-oriented software toolset for probabilistic calibration and prediction in real-world applications.
	
	%
	
	%
	%
	
	
	\bmhead{Supplementary information}
	The article has an accompanying supplementary file.

	\bmhead{Acknowledgements}
	
	This project has received funding from the European Commission under the European Union's Horizon Research and Innovation programme (Grant Agreement No. 101091912 - Project AID4GREENEST) and also from its CLEAN-AVIATION-2022 programme (Grant Agreement No. 101102007 - Project HERA). Views and opinions expressed are, however, those of the authors only and do not necessarily reflect those of the European Union. Neither the European Union nor the granting authority can be held responsible for them.
	
	In addition, CS acknowledges financial support from the Spanish Ministry of Science and Innovation through a Ram\'{o}n y Cajal grant (Grant No. RYC2024-048744-I), financed by MICIU/AEI/10.13039/501100011033 and FSE+.

	\bmhead{Author contributions}
	All authors have made substantial intellectual contributions to the study conception, execution, and design of the work. All authors have read and approved the final manuscript. In addition, the following contributions occurred: 
	Conceptualization: Christina Schenk (CS), Ignacio Romero (IR); Methodology: CS, IR; Formal analysis and investigation: CS, IR; Writing: CS, IR; Original draft preparation: CS, IR; Writing - review and editing: CS, IR; Data Curation: CS, IR; Validation: CS, IR; Visualization: CS; Software: CS, IR; Funding acquisition: CS, IR.
	
	\bmhead{Data and code availability statement}
	All data used in this study were synthetically generated with ACBICI version 2.1.0 and are openly available on GitLab \cite{ACBICIcode}. The complete source code and accompanying documentation are also provided in this repository. The documentation is publicly accessible on ReadtheDocs \cite{ACBICIdocs}.
	
	\section*{Declarations}
	\subsection{Competing interests}
	The authors declare that they have no competing interests.
	\subsection{Ethical Approval}
	Not applicable.
	\bibliography{biblio}
	
	\newpage
	\section*{Supplementary Information}
		\subsection*{Gravity example -  full scripts and complete MCMC and VBMC diagnostics}
	This example illustrates the calibration of a simple physical model that has been 
	previously studied in \cite{mcclarren2018yq}. Specifically, the example shows how
	Bayesian calibration can be employed to infer the probability distribution of a physical constant,
	gravity in this case, from prior information, experimental measurements, and a model. Two
	calibration types are illustrated here: a case $A$ calibration, the simplest of all, and a case $C$
	calibration where the discrepancy is calculated.
	
	In the example, $N$ drop tests are used to estimate the gravitational acceleration $g$. For that, an
	object is dropped from heights $h_i,\ i=1,\ldots,N$ and the times it takes to the floor, indicated as
	$t_i$, are measured and collected in the data set $\mathcal{D}=\{(h_i,t_i), \ i=1,\ldots, N\}$. The
	gravitational acceleration $g$ is the physical constant that, if assumed constant, relates $h$ and
	$t$ through the model
	\begin{equation}
		\label{eq-gravity-t}
		t = \sqrt{\frac{2h}{g}}.
	\end{equation}
	
	To calibrate this model with \ACBICI, it is required to collect the experiments,
	define the model, and select a prior probability distribution for the only unknown parameter,
	namely~$g$. In this example, data are taken from the file \texttt{experiments.dat} and the model is
	defined as a subclass of the \texttt{ACBICImodel} parent class. In the constructor, the dimension of
	the input space ($d=1$), the name of the parameter to be estimated ($g$), and its prior (in this case a uniform
	distribution over $[7,12]$) are given. Then, the analytical expression of the
	model~\eqref{eq-gravity-t} is implemented:
	
	\begin{verbatim}
		class myModel(ACBICImodel):
		def __init__(self):
		self.xdim = 1
		self.addParameter(label=r'$g$', 
		prior=Uniform(a=7.0, b=12.0))
		
		def symbolicModel(self, x, p):
		x = to_column_vector(x)
		p = to_column_vector(p)
		r = np.sqrt(2.0*x[:,0]/p[:,0])
		r = to_column_vector(r)
		return r
	\end{verbatim}
	
	To start the calibration in \ACBICI, the model needs to be instantiated and the
	experimental data is loaded as follows:
	\begin{verbatim}
		theModel = myModel()
		experiments = np.loadtxt("data/experiments.dat")
	\end{verbatim}

	\subsubsection*{Classical calibration with MCMC sampling}\label{ex0A}
	The first calibration is based on type $A$, as described in Section~\ref{ex0A}. In \ACBICI,
	this calibration is performed by a \texttt{classicalCalibrator}, using the following commands.
	First, the calibrator is loaded with the model by instantiating the class, then the the experimental
	data is stored, the calibration process is then executed. Finally, all important plots obtained from
	the analysis are generated.
	\begin{verbatim}
		theCalibrator = classicalCalibrator(theModel)
		theCalibrator.storeExperimentalData(experiments)
		theCalibrator.calibrate()
		theCalibrator.plot()
	\end{verbatim}
	
	The calibration produces a detailed report along with various figures, including autocorrelation,
	calibration, correlation, predictions, and prediction errors. It also provides prior versus
	posterior distributions and trace plots. Regarding the calibration configuration, a Markov Chain
	Monte Carlo (MCMC) sampling method is employed by default using the =emcee= sampler with default
	parameters that can be replaced (see the documentation of \ACBICI\ for all the details on the
	available options). The generated report contains information such as the input and output space
	dimensions, and details about the parameters to be calibrated, including their prior distributions
	and hyperparameters, and their prior distributions. When using MCMC, the report also provides
	information on the chain length, burn-in, number of walkers, the mean acceptance fraction, and the autocorrelation time. 
	
	In addition, the report includes several key diagnostic summaries that help to assess the convergence and reliability of the MCMC sampling process:
	\begin{itemize}
		\item ESS: The report provides both per-parameter ESS and a mean ESS value. ESS quantifies how many independent samples the chain effectively contains after accounting for autocorrelation. The here high ESS values indicate reliable posterior estimates.
		\item $\hat{R}$ Convergence Diagnostic: The split-$\hat{R}$ statistic is reported for each parameter. Values very close to $1$ (here $1.004$) indicate that the chains have mixed well and reached convergence.
		\item Autocorrelation time: The mean autocorrelation time ($\tau$) is included to describe how quickly the chain explores the posterior. The here short $\tau$ of $\approx 36$ steps implies great sampling efficiency.
		\item Posterior sample count: The report clarifies the number of post–burn-in samples (here 128000) and any thinning (here no thinning) applied, providing the final count of posterior samples used in statistical summaries.
		\item Parameter summary table: The summary table lists posterior mean, median, approximate MAP estimate, variance, and 95\% credible intervals. This table provides a concise overview of parameter uncertainty.
		\item Prediction error statistics: The report prints the average, maximum, and standard deviation of calculated prediction errors using the calibrated model, providing an immediate measure of predictive performance.
	\end{itemize}
	
	Additionally, statistical summaries of the parameters -- mean, median, maximum a posteriori (MAP), variance, and 95\% credible intervals -- are provided. Prediction error statistics such as the average and maximum errors, along with the standard deviation of prediction errors, are also summarized.
	
	Beyond these numerical diagnostics, the calibration report serves as a compact record of the entire Bayesian workflow. It documents the prior choices, inference configuration, convergence diagnostics, posterior statistics, and predictive performance, ensuring complete reproducibility and making it straightforward to compare alternative calibration setups in more complex studies.
	
	As indicated above, \ACBICI\ generates when requested figures that illustrate the results of the
	calibration and the process itself.
	These figures include (a) a comparison plot, which visually
	contrasts model predictions with experimental data (cf. Figure~\ref{fig:comparecornerC0a} for the
	example at hand);
	(b) a corner plot such as Figure~\ref{fig:comparecornerA0b}, showing parameter correlations and
	marginal probability distributions;
	(c) a figure with the posterior distribution of the calibrated parameters (cf.
	Figure~\ref{fig:comparecornerA0b} for the posterior distribution of the gravitational acceleration in our example).
	
	Figure~\ref{fig:comparecornerA0a} shows the experimental data (blue markers) alongside the
	predictions from the calibrated model (red line) when the parameter values are given the maximum a
	posterior (MAP) values. When the red line closely follows the blue points, it indicates that the
	calibrated model accurately reproduces the observed data. Good agreement here suggests the
	calibration procedure has successfully identified parameter values that enable the model to fit the
	experimental results. Deviations, if present, highlight regions where the model may have limitations
	or where uncertainty is highest.
	
	Figure~\ref{fig:comparecornerA0b} shows the pairwise joint and marginal posterior distributions of
	the calibrated parameters. The diagonal panels depict the probability density functions for
	individual parameters, with vertical lines for the MAP and mean estimates, as well as dashed lines for the 95\% credible intervals. The off-diagonal panels visualize pairwise parameter correlations. Well-constrained (narrow) marginals and clearly defined credible intervals indicate strong identifiability and certainty about parameter values; significant off-diagonal structure (visible correlations) signals parameter dependencies. 
	Together, these plots validate both the predictive capability of the calibrated model and the uncertainty quantification achieved through Bayesian inference, providing a robust foundation for subsequent analysis or model use.
	\begin{figure}[htb]
		\centering
		\includegraphics[width=.85\linewidth]{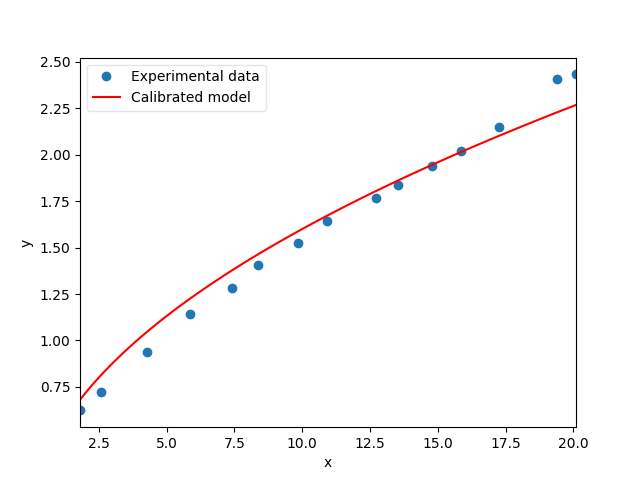}
		\caption{Example~\ref{ex0A}. Comparison between experimental data (blue markers) and calibrated model predictions (red line). Close agreement indicates accurate reproduction of the observed behavior using the MAP parameter estimates.}
		\label{fig:comparecornerA0a}
	\end{figure}
	\begin{figure}[htb]
		\centering
		\includegraphics[width=.85\linewidth]{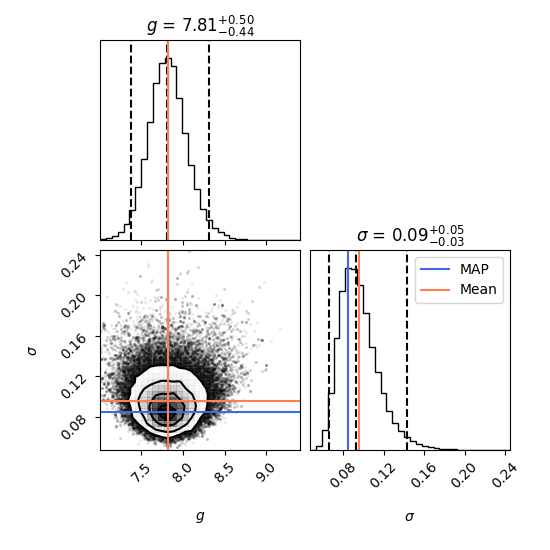}
		\caption{Example~\ref{ex0A}. Corner plot showing marginal and joint posterior distributions of the calibrated parameters. Diagonal panels display marginal densities with MAP, mean, median, and 95\% credible intervals; off-diagonal panels illustrate parameter correlations.}
		\label{fig:comparecornerA0b}
	\end{figure}
	
	For this example, the diagnostics demonstrate, overall, a well-converged MCMC calibration. The
	posterior distributions are sharply concentrated relative to the priors, indicating strong
	information gain from the data. Convergence metrics such as ESS and $\hat{R}$ confirm excellent
	chain mixing and sampling reliability. The tight credible intervals for both $g$ and $\sigma$, along
	with low prediction errors, show that the calibrated model accurately reproduces the observed data
	and yields meaningful uncertainty quantification. This example serves as a first validation of the
	calibration pipeline and illustrates how \ACBICI\ reports and figures provide all necessary
	information for both diagnostic assessment and model interpretation.
	
	\subsubsection*{Discrepancy calibration with variational Bayesian Monte Carlo}\label{ex0C}
	We extend the previous calibration method by accounting now for possible structural inadequacies in the model. To represent such mismatch explicitly, we employ the \texttt{discrepancyCalibrator} class with variational Bayesian Monte Carlo. The model discrepancy is modeled using Gaussian process surrogates.
	\begin{verbatim}
		theCalibrator = discrepancyCalibrator(theModel, name="typeC")
		theCalibrator.storeExperimentalData(experiments)
		theCalibrator.calibrate(method="vbmc")
		theCalibrator.plot()
	\end{verbatim}
	Running the discrepancy-aware calibration using the \texttt{discrepancyCalibrator} class with selecting variational Bayes as our method of choice through method\ ="vbmc", results in the following outcomes (see Figures~\ref{fig:comparecornerC0a},\ref{fig:comparecornerC0b} and \ref{fig:discrepancyCplot}). 
	\begin{figure}[htb]
		\centering
		\includegraphics[width=0.65\linewidth]{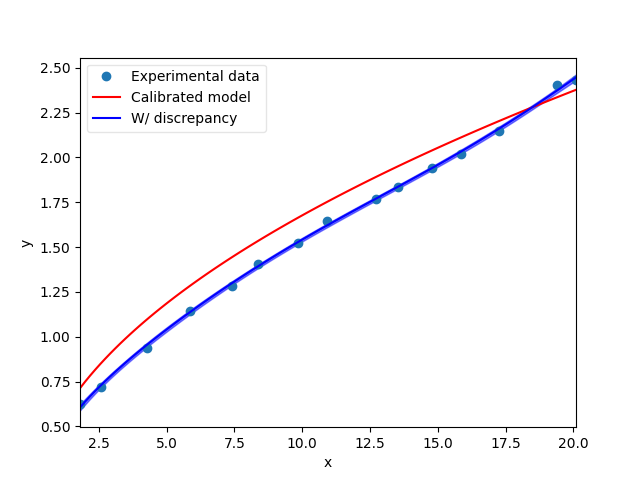}
		\caption{Example~\ref{ex0C}. Comparison between experimental data (markers) and model predictions after discrepancy calibration. The discrepancy-corrected predictions provide improved agreement with the observations, particularly in regions where the base model exhibits systematic deviations.}
		\label{fig:comparecornerC0a}
	\end{figure}
	
	\begin{figure}[htb]
		\centering
		\includegraphics[width=0.85\linewidth]{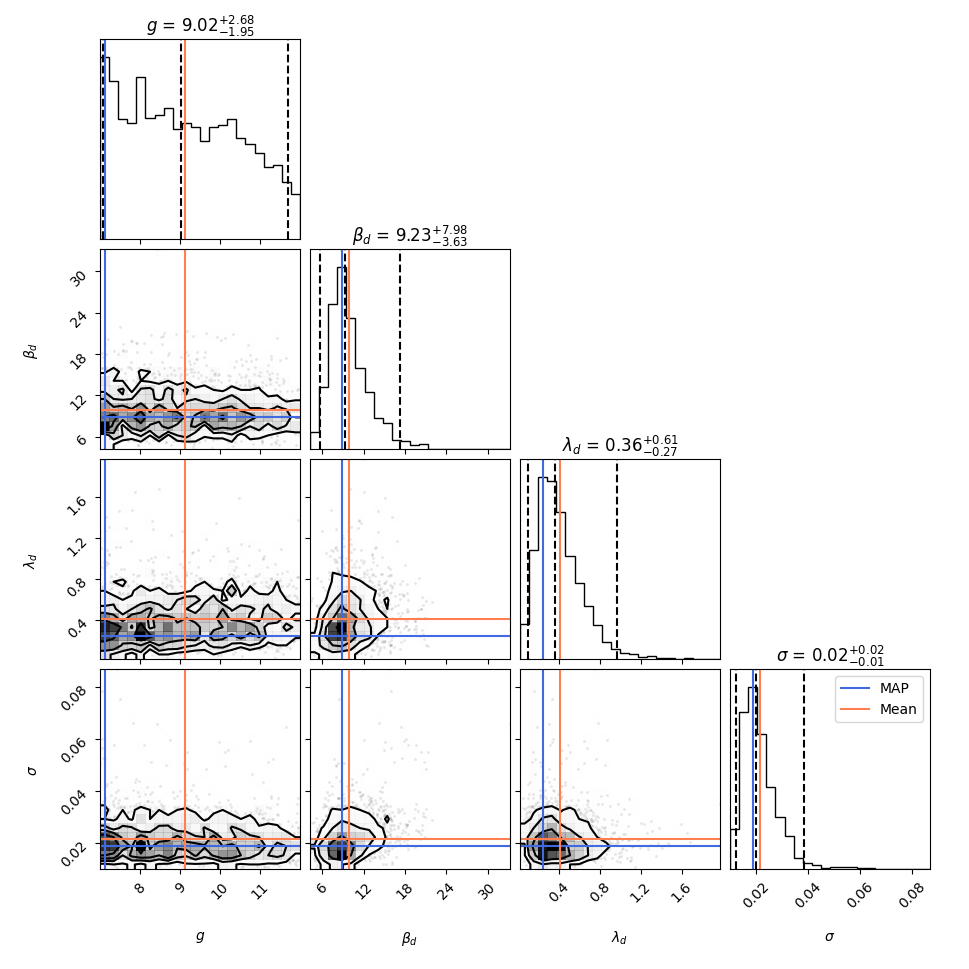}
		\caption{Example~\ref{ex0C}. Posterior distributions and parameter correlations obtained via variational Bayes. The widened distribution for \(g\) reflects the uncertainty introduced by explicitly modeling structural discrepancy, while the additional hyperparameters characterize the inferred discrepancy process.}
		\label{fig:comparecornerC0b}
	\end{figure}
	\begin{figure}[htb]
		\centering
		\includegraphics[width=0.65\linewidth]{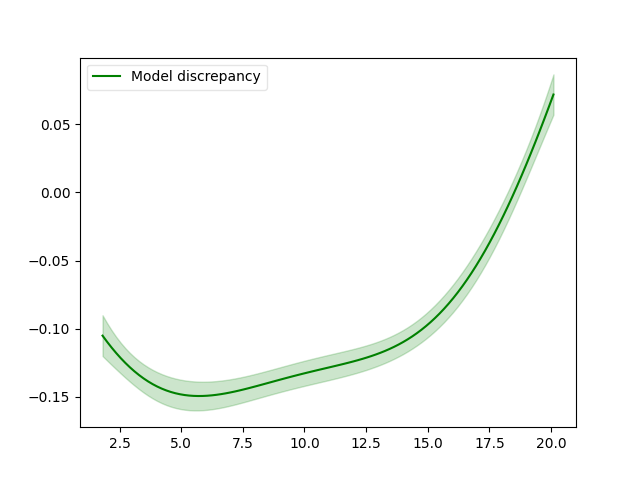}
		\caption{Example~\ref{ex0C}. Estimated model discrepancy using a Gaussian process surrogate, reflecting systematic deviations between the model and data.}
		\label{fig:discrepancyCplot}
	\end{figure}
	Figure~\ref{fig:comparecornerC0a} shows the experimental observations (blue dots), the predictions from the calibrated model without discrepancy (red line), and the predictions accounting for the inferred model discrepancy (blue line). The calibrated model alone fits well, but the addition of the discrepancy correction brings predictions even closer to the experimental data, particularly at the boundaries and in regions where systematic deviation is present. This demonstrates the benefit of explicitly modeling discrepancy to capture effects not explained by the original model structure.
	The corner plot Figure~\ref{fig:comparecornerC0b} displays joint and marginal posterior distributions for all parameters, including , under discrepancy-aware calibration. The posterior for  is notably broader than in classical (discrepancy-free) calibration, with a MAP value of ~ $7.796$, a mean value of ~$9.197$, a median value of $9.043$ and a credible interval that spans a wider range. This reflects reduced certainty about $g$ after accounting for discrepancy, which is expected: since the discrepancy term absorbs systematic mismatch, the data become less informative about the exact value of $g$. The broader posterior thus prevents overconfident inference about $g$, leading to more honest uncertainty quantification.
	The inferred discrepancy function (cf. Figure~\ref{fig:discrepancyCplot} captures systematic trends missed by the calibrated model, and its credible band quantifies uncertainty in these corrections.
	
	To ensure convergence, we analyze the diagnostics reported in the log file. Across iterations, the estimated evidence lower bound (ELBO) increased rapidly during the warm-up phase and then plateaued, with the ELBO standard deviation shrinking to small values, indicating a stable and well-resolved objective. In parallel, the symmetrized Kullback-Leibler (KL) divergence between consecutive variational posteriors (sKL-iter) decreased to near zero, showing that the approximate posterior stopped changing appreciably. These signals are combined by PyVBMC into a reliability/convergence index (r\_index; reported as “Convergence”), which fell below the recommended threshold ($<1$) and remained low for the required number of function evaluations. Thus, the algorithm terminated with the status ``variational solution stable'' and a final ELBO estimate reported with tight uncertainty.
	
	In addition, the calibration report provides information on the convergence of the variational Bayes optimization (e.g., ELBO evolution), and, in the context of surrogate-based discrepancy estimation, reports on the selected Gaussian process kernel.
	
	Beyond these numerical summaries, the discrepancy-aware calibration report includes several diagnostic elements that help interpret how the model–data gap is handled.
	First, the kernel information confirms that a squared exponential Gaussian process is used, implying smooth discrepancy corrections. The calibrated hyperparameters $\beta_d$ and $\lambda_d$ show that the discrepancy has nontrivial amplitude and a moderate characteristic length scale, allowing it to absorb systematic deviations in the experiment without overfitting noise.
	
	Second, the posterior for the physical parameter $g$ is significantly broader than in the classical calibration, with a 95\% credible interval of $[7.067, 11.70]$. This widening reflects that the discrepancy accounts for structural mismatch, reducing the need for $g$ to compensate for bias in the model. As a result, the inference becomes less precise but more realistic.
	
	Third, the noise parameter $\sigma$ contracts toward a smaller value (mean value $\approx 0.022$), indicating that once model discrepancy is accounted for, the remaining residuals are interpreted as genuine measurement noise rather than unmodeled physics.
	
	The prediction error statistics show an average error of $0.11$ and a relatively small standard deviation, reflecting a balance between fitting the data and representing structural uncertainty. The optimization trajectory reported by PyVBMC (ELBO increasing, KL divergence decreasing, final convergence index $< 0.3$) confirms that the variational solution is stable and trustworthy.
	
	Overall, discrepancy-aware Bayesian calibration yields wider, more realistic credible intervals for physical parameters such as $g$, reflecting both parametric and structural uncertainties. This results in posterior inferences that are less precise but more trustworthy, and improves the reliability of predictions in the presence of potential model inadequacy.
	
	\subsection*{Cobb–Douglas example – full scripts, diagnostic plots, and numerical summaries} \label{ex3}
	This example utilizes the Cobb-Douglas production function, a widely recognized model in economics, to relate total output \(Y\) to inputs of labor \(L\), capital \(K\), and a productivity factor \(T\). The model is expressed as:
	
	\begin{equation}
		Y = T \cdot L^{\alpha} \cdot K^{\gamma},
	\end{equation}
	
	where \(\alpha\) and \(\gamma\) are parameters indicating the elasticities of output with respect to labor and capital, respectively.
	
	\subsubsection*{Input scaling}
	\label{sec:scaling}
	
	To ensure numerical stability during calibration, inputs are nondimensionalized. The parameter \(T\) is assumed to be within \([0, 1]\). The physical inputs \(L\) and \(K\) are scaled using reference values \(L_0 = 50\) and \(K_0 = 200\):
	
	\begin{align*}
		\tilde{L} &= \frac{L}{L_0}, \\
		\tilde{K} &= \frac{K}{K_0}.
	\end{align*}
	
	The resulting scaled input vector used in the model becomes:
	
	\begin{equation}
		x = (T, \tilde{L}, \tilde{K}).
	\end{equation}
	
	\subsubsection*{Model definition}
	The model is implemented by creating a subclass of \texttt{ACBICImodel} as shown below:
	
	\begin{verbatim}
		class myModel(ACBICImodel):
		
		def __init__(self):
		self.xdim = 3
		self.addParameter(label=r'$\alpha$', 
		prior=Uniform(a=0.3, b=1.6))
		self.addParameter(label=r'$\gamma$', 
		prior=Normal(mu=0.35, sigma=0.1))
		
		def symbolicModel(self, x, p):
		T = x[:, 0]
		L = x[:, 1]
		K = x[:, 2]
		alpha = p[:, 0]
		gamma = p[:, 1]
		r = T * L**alpha * K**gamma
		return r.reshape(-1, 1)
	\end{verbatim}
	\subsubsection*{Calibration procedure}
	Calibration proceeds with the same methodology as for previous examples, loading the calibrator with the model and storing the data and then running calibration. Here, however, we consider the case in which the Cobb–Douglas model is treated as computationally expensive, and we therefore rely on a Gaussian process surrogate during sampling. We also estimate the experimental error simultaneously using a Gamma prior. Calibration is performed via the \texttt{expensiveCalibrator}, i.e. 
	\begin{verbatim}
		print("*** Type B calibration + unknown experimental error")
		calibratorBexp = expensiveCalibrator(theModel, name="calBE")
		calibratorBexp.setExperimentalSTDPrior(Gamma(alpha=2,beta=2))
		calibratorBexp.storeExperimentalData(experiments)
		calibratorBexp.calibrate(nsteps=15000)
		calibratorBexp.plot(onscreen=False)
	\end{verbatim}
	The default calibration configuration is retained except for an increased number of steps nsteps=15000 The prior on the standard deviation of the experimental error is set as a gamma distribution with shape parameter $\alpha=2$ and rate parameter $\beta=2$.
	
	\begin{figure}[htb]
		\centering
		\includegraphics[width=0.65\linewidth]{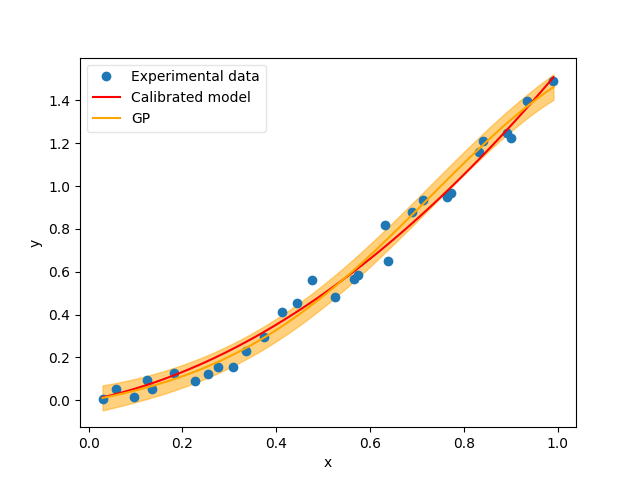}
		\caption{Example \ref{ex3}. Comparison of experimental data for the first input dimension (markers), calibrated model predictions (red), and uncertainty band for the GP-based surrogate (orange). The model and surrogate closely follow the observed trend, indicating successful calibration and accurate surrogate representation.}
		\label{fig:compare3a}
	\end{figure}
	
	\begin{figure}[htb]
		\centering
		\includegraphics[width=0.85\linewidth]{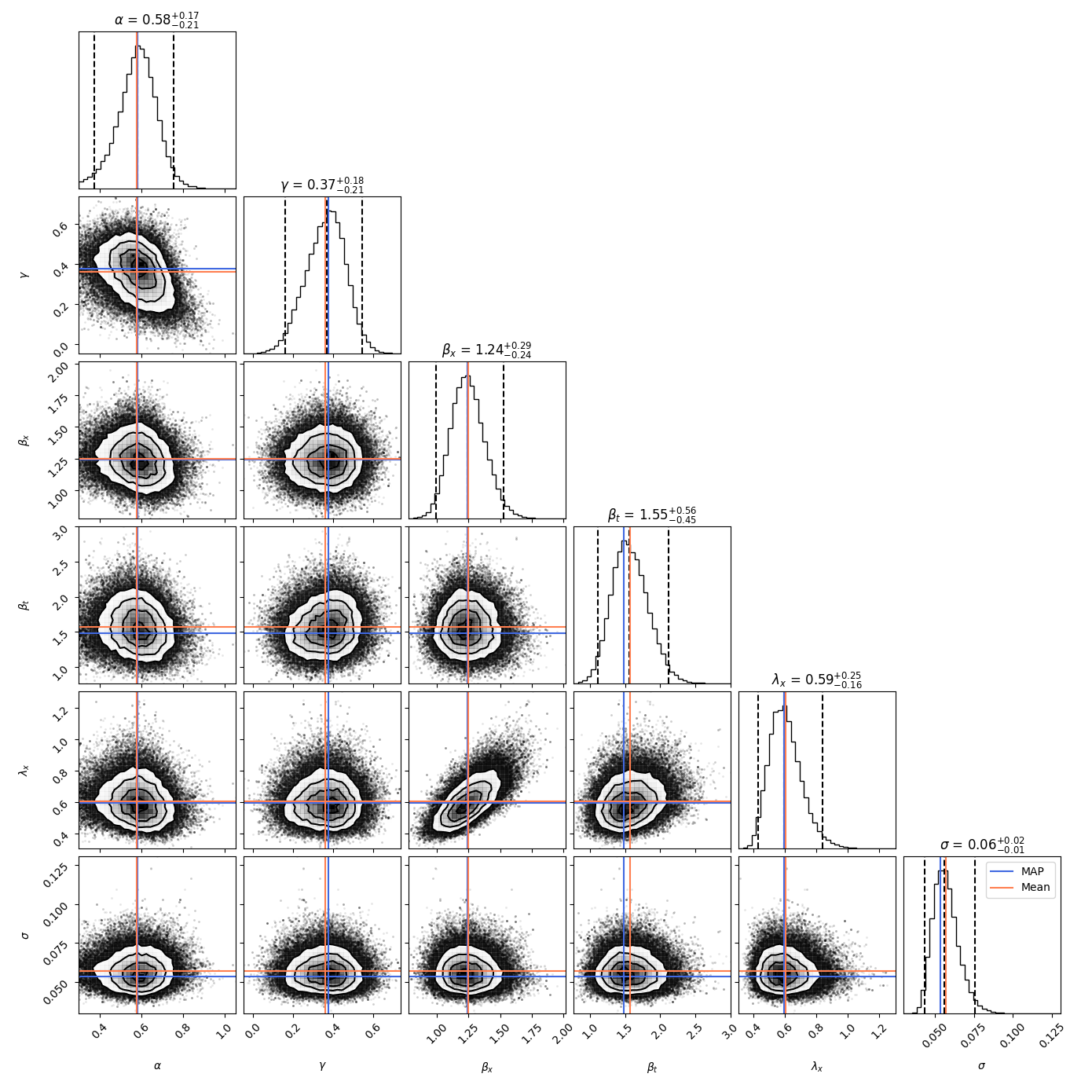}
		\caption{Example \ref{ex3}. Posterior distributions and pairwise correlations for the calibrated parameters \(\alpha\), \(\gamma\), and the GP hyperparameters. Marginal distributions show credible intervals and parameter uncertainty, while joint distributions reveal dependencies.}
		\label{fig:corner3b}
	\end{figure}
	The model fit in Figure~\ref{fig:compare3a} demonstrates strong agreement between calibrated predictions and experimental data, with the GP surrogate effectively capturing residual structure and uncertainty. Figure~\ref{fig:corner3b} shows posterior distributions for model and GP hyperparameters, quantifying both identifiability and uncertainty within the calibration. Detailed calibration diagnostics, including autocorrelation, summary statistics, and prediction error analysis, are reported in the output log and other graphical outputs. 
	
	The calibrator information indicates that a squared exponential kernel was selected for the GP surrogate, implying that the underlying response surface is expected to vary smoothly with respect to T, $\tilde{L}$, and $\tilde{K}$. The three GP hyperparameters $\beta_x$, $\beta_t$, and $\lambda_x$ control the overall amplitude and characteristic length scales of the surrogate. Their posterior means ($\approx 1.24, 1.57,$ and $0.60$) show moderate smoothing and demonstrate that the surrogate adapts flexibly across different input directions.
	
	The MCMC performance metrics in the process report confirm a well-behaved and converged calibration. The mean acceptance fraction (0.510), the mean moderate autocorrelation time ($75$ steps) and the very high ESS ($\approx 2331–2722$ per parameter) reflect efficient sampling, while split-$\hat{R}$ values near $1.0$ confirm convergence.
	
	The posterior summaries reveal that both production elasticities are well-identified from the data. The posterior for $\alpha$ centers near $0.58$ with a 95\% credible interval $[0.376, 0.754]$. Similarly, $\gamma$ concentrates near $0.36$, with uncertainty reflecting the sensitivity of output to capital. These results demonstrate that the calibration procedure successfully recovers plausible elasticity estimates while properly quantifying uncertainty.
	
	The GP hyperparameters $\beta_x$, $\beta_t$, and $\lambda_x$ exhibit tight credible intervals, indicating that the surrogate model is well informed by the data and not overfitting. The posterior for the experimental noise level $\sigma$ is also sharply concentrated ($\approx 0.057$), showing that measurement error is relatively small once model and surrogate uncertainties are accounted for.
	
	Prediction error statistics further validate the calibration: the average prediction error is slightly below $0.046$, with a maximum error around $0.11$ and a small standard deviation. These values indicate that the calibrated model and surrogate jointly capture the input–output relationship accurately and consistently across the dataset.
	
	Overall, the calibration results show strong parameter identifiability, stable MCMC convergence, an informative surrogate model, and excellent predictive performance. This example demonstrates how \ACBICI\ can be used to calibrate nonlinear, economically meaningful models even in cases where the model is treated as computationally expensive.
	
	\subsection*{Traction test example - full scripts, multi-output diagnostics, and all hyperparameter/posterior tables}\label{ex4}
	
	\begin{figure}[htb]
		\centering
		\begin{tikzpicture}[line join=round, line cap=round, >=Latex]
			\def\L{6.0}     
			\def\H{1.3}     
			\def\D{1.0}     
			\def\shift{0.9} 
			
			\coordinate (FBL) at (\L,0);          
			\coordinate (FBR) at (\L+\D,0.4);     
			\coordinate (FTR) at (\L+\D,\H+0.4);  
			\coordinate (FTL) at (\L,\H);         
			
			\coordinate (BBL) at (0,0);
			\coordinate (BBR) at (\D,0.4);
			\coordinate (BTR) at (\D,\H+0.4);
			\coordinate (BTL) at (0,\H);
			
			\pgfmathsetmacro{\Hwall}{\H+0.8}
			\pgfmathsetmacro{\Hhatch}{\H+0.6}
			
			\draw[thick] (0.0,0.3) -- (0.0,\Hwall);
			
			\foreach \y in {-0.2,0.1,...,\Hhatch}{
				\pgfmathsetmacro{\yshift}{\y+0.4}
				\draw[thick] (-.3,\y) -- (0.,\yshift);
			}
			
			\draw[thick] (BBL) -- (FBL) -- (FTL) -- (BTL) -- cycle;  
			
			\draw[thick] (FBR) -- (FTR) -- (BTR);
			
			\draw[thick] (FBL) -- (FBR);  
			\draw[thick] (FTL) -- (FTR);  
			\draw[thick] (BTL) -- (BTR);  
			
			\fill[gray!8] (FBL) -- (FBR) -- (FTR) -- (FTL) -- cycle;
			\node[font=\large] at ($(FBL)!0.38!(FTR)$) {$A$};
			
			\draw[very thick, -{Latex[length=4mm]}]
			($(FBL)!0.5!(FTL) + (0.5,0.2)$) -- ++(3.0,0.0)
			node[right, font=\large] {$F$};
			
			\tikzset{hidden/.style={thick,dashed,gray!70}}
			
			\draw[hidden] (BBL) -- (BBR);
			\draw[hidden] (BBR) -- (BTR);
			\draw[hidden] (BBR) -- (FBR);
			\draw[thick] (0,-0.9) -- (0,-0.2);
			\draw[thick] (\L,-0.9) -- (\L,-0.2);
			
			\draw[thick, <->]
			(0,-0.75) -- (\L,-0.75)
			node[midway, fill=white, inner sep=1pt, font=\large] {$L$};
			
		\end{tikzpicture}
		\caption{Traction test schematic.}
		\label{fig:ex4sketch}
	\end{figure}
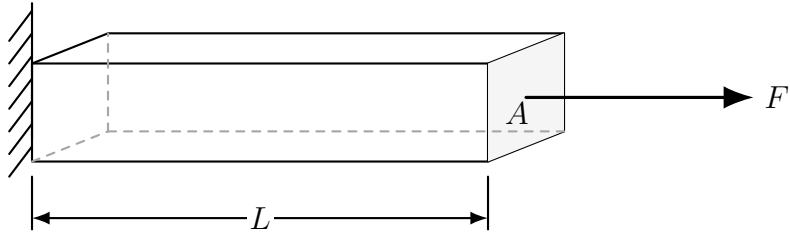
	
	We now consider a traction-test example to demonstrate how \ACBICI\ handles multi-output Bayesian calibration, where two coupled physical responses are inferred simultaneously from experimental data. This traction-test example for an elastic isotropic material is sketched in Figure~\ref{fig:ex4sketch}. The physical relations under an applied traction force \(F\) are:
	
	\begin{equation}
		y_1 = L \left( 1 + \frac{F}{A E} \right), \quad
		y_2 = A \left( 1 - \frac{F \nu}{A E} \right)^2,
	\end{equation}
	
	where \(L\) and \(A\) are the specimen's initial length and cross-sectional area, \(E\) the Young's modulus, and \(\nu\) the Poisson's ratio.
	
	\subsubsection*{Nondimensionalization}
	
	To ensure consistency and comparability across outputs in the multi-objective calibration, variables are nondimensionalized by characteristic references:
	
	\begin{equation}
		\hat{F} = \frac{F}{F_0}, \quad
		\hat{E} = \frac{E}{E_0}, \quad
		\hat{L} = \frac{L}{L_0}, \quad
		\hat{A} = \frac{A}{A_0},
	\end{equation}
	
	with \(E_0 = 250\,\text{MPa}\), \(L_0 = 10\,\text{mm}\), \(A_0 = 1\,\text{mm}^2\), and \(F_0 = E_0 \times A_0 = 250\,\text{N}\).
	
	The nondimensional outputs are thus:
	
	\begin{equation}
		y_1 = \hat{L} \left(1 + \frac{\hat{F}}{\hat{A} \hat{E}}\right), \quad
		y_2 = \hat{A} \left(1 - \frac{\hat{F} \nu}{\hat{A} \hat{E}}\right)^2,
	\end{equation}
	
	and for the given geometry \(\hat{L} = \hat{A} = 1\).
	
	\subsubsection*{Model definition}
	
	The model is defined by subclassing \texttt{ACBICImodel} and specifying \(\texttt{ydim} = 2\) for two simultaneous outputs:
	
	\begin{verbatim}
		class myModel_nondim(ACBICImodel):
		
		def __init__(self):
		self.xdim = 1
		self.ydim = 2
		E0 = 250.
		self.addParameter(label=r'$E$', 
		prior=Uniform(a=200.0/E0, b=400.0/E0))
		self.addParameter(label=r'$\nu$', 
		prior=Uniform(a=0.0, b=0.5))
		
		def symbolicModel(self, x, p):
		x = np.atleast_2d(x)
		p = np.atleast_2d(p)
		
		L = 10.
		b = 1
		h = 1
		A = b * h
		A0 = A
		L0 = 10
		A_hat = A/A0
		L_hat = L/L0
		
		r1 = L_hat * (1 + x[:, 0] / (A * p[:,0])) 
		# length response
		r2 = A_hat * (1 - x[:, 0] * p[:,1] / (A * p[:,0]))**2 
		# area response
		
		r = np.column_stack([r1, r2])
		return r
	\end{verbatim}
	
	\subsubsection*{Calibration procedure}
	The multi-output calibration works for all calibration types -- classical, discrepancy-aware, surrogate-based, and combined. Here, an exemplary surrogate-based and discrepancy-aware calibration is performed using the \texttt{KOHCalibrator}. Unknown experimental noise is assumed and the following configuration parameters are set: thin=1, burn=0.2, nwalkers=16, nsteps=20000.
	
	\begin{verbatim}
		print("*** Type D calibration + unknown experimental error 
		+ external synthetic")
		theCalibrator = KOHCalibrator(theModel, name="calDE")
		theCalibrator.storeExperimentalData(experiments)
		theCalibrator.storeSyntheticData(sd)
		theCalibrator.calibrate(thin=1, burn=0.2, nwalkers=16, 
		nsteps=20000)
		theCalibrator.plot(dumpfiles=True, onscreen=False)
	\end{verbatim}
	Both outputs are fitted simultaneously, and the calibration procedure quantifies both prediction errors and parameter uncertainties.
	The results are highlighted in Figures~\ref{fig:predictions4a} and \ref{fig:corner4b}:
	\begin{figure}[htb]
		\centering
		\includegraphics[width=0.65\linewidth]{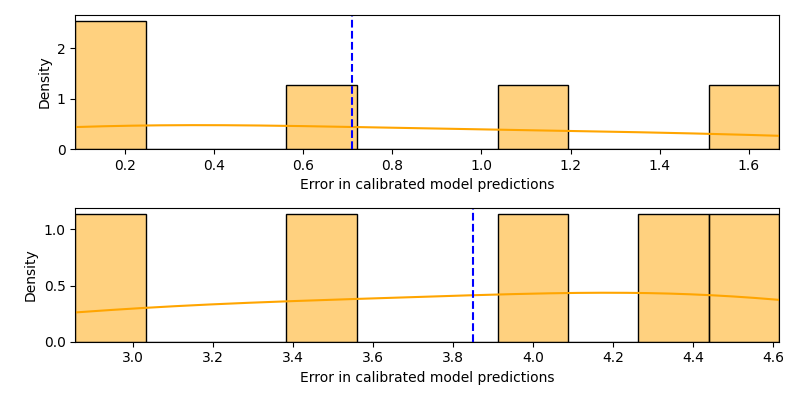}
		\caption{Example~\ref{ex4}. Prediction error plot showing residuals for both outputs across the input range. Clustering of residuals near zero indicates good predictive performance, while broader spreads highlight regions of increased uncertainty or model mismatch.}
		\label{fig:predictions4a}
	\end{figure}
	
	\begin{figure}[htb]
		\centering
		\includegraphics[width=0.85\linewidth]{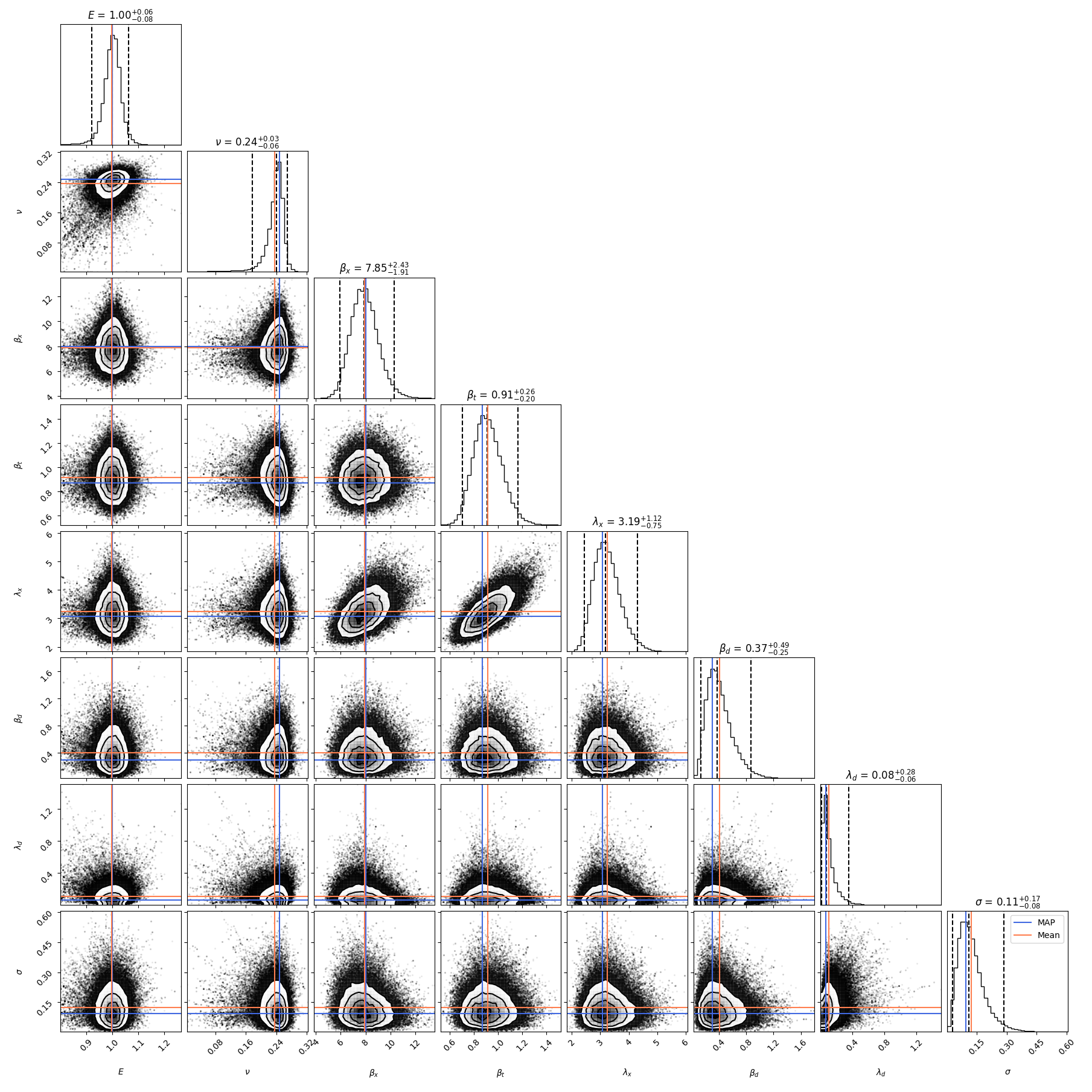}
		\caption{Example~\ref{ex4}. Posterior distributions and pairwise correlations for all calibrated and surrogate-related parameters. Marginal distributions provide credible intervals and uncertainty quantification, while joint distributions reveal parameter dependencies and identifiability.}
		\label{fig:corner4b}
	\end{figure}
	The prediction error plot (Figure~\ref{fig:predictions4a}) illustrates the density and spread of residuals for each output individually, providing detailed insight into the model's predictive accuracy across multiple objectives. Residuals clustered near zero indicate that the calibrated model generally makes accurate predictions for both outputs. Since the reported prediction error statistics -- average error \(2.17\), maximum error \(5.50\), and standard deviation \(1.91\) -- are averaged over all outputs, they represent an aggregated measure of predictive performance. This averaging can mask output-specific behaviors, highlighting the importance of examining residual distributions separately for each output (as in Figure~\ref{fig:predictions4a}) when assessing multi-output calibration quality.
	
	The shape and spread of residual distributions reflect both the inherent randomness in the data and the uncertainty quantified through calibration. Some outputs may exhibit wider residual spreads or biases that are averaged out in global statistics but are critical for understanding model limitations in specific outputs or regimes. The presence of larger errors (indicated by the maximum) suggests that while the model performs well overall, there exist instances where predictions deviate significantly from observations.
	In summary, combining averaged prediction error statistics with detailed residual density plots for each output offers a comprehensive approach to assessing multi-objective calibration quality, providing both a global performance summary and granular insights into output-specific predictive reliability.
	
	The corner plot (Figure~\ref{fig:corner4b}) displays the joint and marginal posterior distributions for all model and surrogate parameters. Marginal distributions provide credible intervals, means, and MAP estimates, while the off-diagonal scatter plots reveal parameter correlations and identifiability. The principal model parameters (\(E\) and \(\nu\)) appear well-constrained, with substantive separation between surrogate parameters. Together, these results demonstrate robust parameter inference and successful multi-objective calibration, with credible quantification of predictive error and uncertainty.
	
	\begin{figure}[htb]
		\centering
		\includegraphics[width=0.75\linewidth]{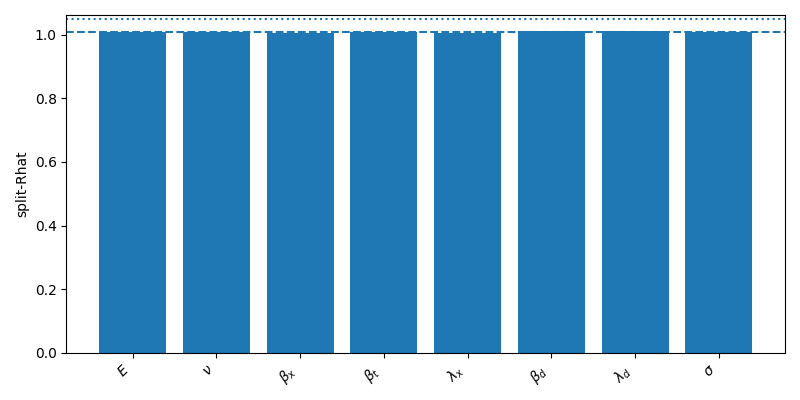}
		\caption{Example \ref{ex4}. Split-\(\hat{R}\) convergence diagnostics for all calibrated and surrogate-related parameters.
			Bars correspond to individual parameters (\(E\), \(\nu\), \(\beta_x\), \(\beta_t\), \(\lambda_x\), \(\beta_d\), \(\lambda_d\), and \(\sigma\)),
			while the two dashed lines mark the recommended convergence thresholds of $1.01$ and $1.05$.
			Values very close to 1 indicate that all chains have mixed well and that no parameter shows signs of non-convergence.}
		\label{fig:rhat4}
	\end{figure}
	In addition to the prediction and posterior plots, Figure~\ref{fig:rhat4} summarizes the split-\(\hat{R}\) convergence diagnostics for all calibrated and surrogate-related parameters. All values lie extremely close to one and below both displayed threshold lines, indicating that no parameter exhibits problematic convergence behavior and that the MCMC chains have mixed satisfactorily.
	
	As in previous examples, detailed diagnostics -- including autocorrelation, prediction and calibration error statistics, and prior versus posterior assessment -- are provided in the output report and graphical summaries.
	
	The calibrator information indicates that a nonstationary multi-task kernel is employed, enabling the Gaussian process surrogate to model both outputs jointly while allowing for output-specific behavior. The hyperparameters \(\beta_x\), \(\beta_t\), \(\lambda_x\), \(\beta_d\), and \(\lambda_d\) control, respectively, the overall amplitude and length scales in the input space, as well as the structure and smoothness of the discrepancy component. Their posterior means and relatively narrow credible intervals show that the surrogate and discrepancy terms are well informed by the data and do not exhibit implausible behavior.
	
	The MCMC performance metrics confirm a stable and convergent calibration run. The mean acceptance fraction of $0.429$, the mean autocorrelation time of about 142 steps, combined with an ESS on the order of \(10^3\) per parameter, indicate satisfactory chain mixing and a large number of effectively independent samples. Split-\(\hat{R}\) values between $1.007$ and $1.012$ (cf.~Figure~\ref{fig:rhat4}) are very close to unity, providing strong evidence of convergence across all parameters.
	
	The posterior summaries show that the nondimensional Young's modulus \(E\) concentrates tightly around unity (\(\mathbb{E}[E]\approx 0.998\)), consistent with the chosen reference \(E_0\), and that the Poisson's ratio \(\nu\) is well identified with mean \(\approx 0.24\) and MAP \(\approx 0.25\) and a relatively narrow 95\% credible interval. The surrogate and discrepancy hyperparameters display physically reasonable ranges, reflecting smooth but flexible corrections, while the experimental noise level \(\sigma\) is moderately small and well constrained. Overall, these diagnostics indicate that the multi-output KOH calibration has produced a well-resolved posterior, with reliable uncertainty quantification for both material parameters and surrogate components.
	
	Overall, this example demonstrates that \ACBICI\ can reliably calibrate multi-output mechanical models, yielding well-resolved posteriors and accurate predictive performance across multiple response quantities.
	
	\subsection*{ACBICI distribution and installation}
	While Unix-based operating systems are recommended for optimal performance and compatibility, the software is also functional on Windows platforms when used within an Anaconda environment, broadening accessibility across diverse computing setups. This cross-platform flexibility ensures researchers can effectively employ \ACBICI\ in various environments without compromising usability.
	
	With a focus on user-friendliness, the software includes a basic graphical user interface (GUI) that provides default settings optimized for common calibration tasks. This design aims to lower entry barriers for new users by simplifying workflow setup. 
	
	
\end{document}